\documentclass[superscriptaddress,aps,pra,twocolumn,preprintnumbers,amsmath,amssymb,floatfix,showpacs]{revtex4-2}

\usepackage{graphicx}
\usepackage{dcolumn}
\usepackage{bm}
\usepackage{amssymb}
\usepackage{subfigure}
\usepackage{xcolor}
\usepackage{stmaryrd}
\usepackage[all]{xy}
\usepackage{textcomp}
\usepackage{xr}

\newcommand{\bra}[1]{\ensuremath{\left\langle#1\right|}}
\newcommand{\ket}[1]{\ensuremath{\left|#1\right\rangle}}
\newcommand{\braket}[2]{\left\langle #1\middle|#2\right\rangle}

\begin{document}

\title{Optical and spin manipulation of non-Kramers rare-earth ions under weak magnetic field for quantum memory applications}

\author{J. Etesse}

\affiliation{Universit\'e c\^ote d'Azur, CNRS, Institut de Physique de Nice (INPHYNI), UMR 7010, Parc Valrose, Nice Cedex 2, France}

\author{A. Holz\"apfel}
\author{A. Ortu}
\author{M. Afzelius}
\affiliation{Department of Applied Physics, University of Geneva, CH-1211 Geneva 4, Switzerland}

\date{\today}


\begin{abstract}
Rare-earth ion doped crystals have proven to be solid platforms for implementing quantum memories. Their potential use for integrated photonics with large multiplexing capability and unprecedented coherence times is at the core of their attractiveness. The best performances of these ions are however usually obtained when subject to a dc magnetic field, but consequences of such fields on the quantum memory protocols have only received little attention. In this article, we focus on the effect of a dc bias magnetic field on the population manipulation of non-Kramers ions with nuclear quadrupole states, both in the spin and optical domains, by developing a simple theoretical model. We apply this model to explain experimental observations in a ${}^{151}$Eu:Y$_2$SiO$_5$ crystal, and highlight specific consequences on the AFC spin-wave protocol. The developed analysis should allow to predict optimal magnetic field configurations for various protocols.
\end{abstract}


\maketitle

\section*{Introduction}

Processing and distribution of quantum information has seen tremendous progress in the recent years, thanks to the development of architectures for quantum computing \cite{Arute_2019} and deployment of photonics-based large-scale quantum networks \cite{Boaron18,Liao_2017}. A crucial ingredient is however still missing in order to fully synchronize the elementary photonic links in the networks, and implement for instance quantum repeaters \cite{Sangouard11}: the quantum memories \cite{Heshami16}. Different platforms exist for implementing these devices \cite{Radnaev2010,Nicolas14,Yang2016,Pu2017,Tian2017}, but a particularly interesting one has emerged over the last decade for this purpose: the rare-earth ion doped crystals (REIDC) \cite{Longdell2005,Usmani10,Heinze13,Ferguson2016,Seri2017,Laplane2017,Holzapfel20}. Together with the spin-wave atomic frequency comb (AFC) protocol \cite{Afzelius09}, state-of-the art performances have been demonstrated with these elements in terms of storage duration \cite{Heinze13,Zhong15,Holzapfel20}, multiplexing capacity \cite{Usmani10,Sinclair2014}, ability to store single photons \cite{Ferguson2016,Laplane2017,Seri2017} and potential for high-efficiency storage \cite{Sabooni2013,Jobez2014}. A fully favorable regime for operating REIDC is usually at magnetic fields that allow to reach a point where the magnetic sensitivity to environmental fluctuations is minimal (so-called Zero First-Order Zeeman (ZEFOZ) points) \cite{Fraval2004,Heinze13,Longdell2006}. However this situation requires a higher degree of control: magnetic field amplitudes could be high, and their direction has to be precisely adjusted \cite{Zhong15}. On the other hand, working with intermediate amplitude magnetic fields ($\sim$10 to 100 mT) also results in larger effective coherence times \cite{Equall1994} and allows to reach storage times of the order of a second \cite{Holzapfel20}. So far, only a few studies have been conducted to identify side-effects of such a field on the protocol at these field intensities, and experimental requirements are still widely unknown. In this article, we identify phenomena affecting the performances of population manipulations both in the optical and in the spin domain for REIDC, and identify consequences on the spin-wave AFC protocol in this regime.
\\
The paper is organized as follows: in section I, we recall the principle of the spin-wave AFC protocol and its application to REIDC, and describe the different experimental setups used for the experiments of this paper. In section II, we focus on the influence of an external bias field on the spin level manipulations. The studies are conducted in the case of a 3/2 nuclear spin, governed by an effective quadrupole Hamiltonian, and theoretical findings are confronted to experimental observations. In section III, the influence of the field on the population preparation necessary for the AFC protocol is addressed, and the temporal dependance of the AFC spin-wave efficiency under such field is closely studied. In particular, we identify frequencies that appear in the efficiency curve, and associate them with beatings between different quantum paths linked to the lift of the Zeeman degeneracy.\medskip

\section{Spin-wave AFC in REIDC}
\label{sectionNoField}
\subsection{The AFC protocol}
\label{sub_AFCprot}
\begin{figure}[!h]
\begin{center}
\includegraphics[width=7cm]{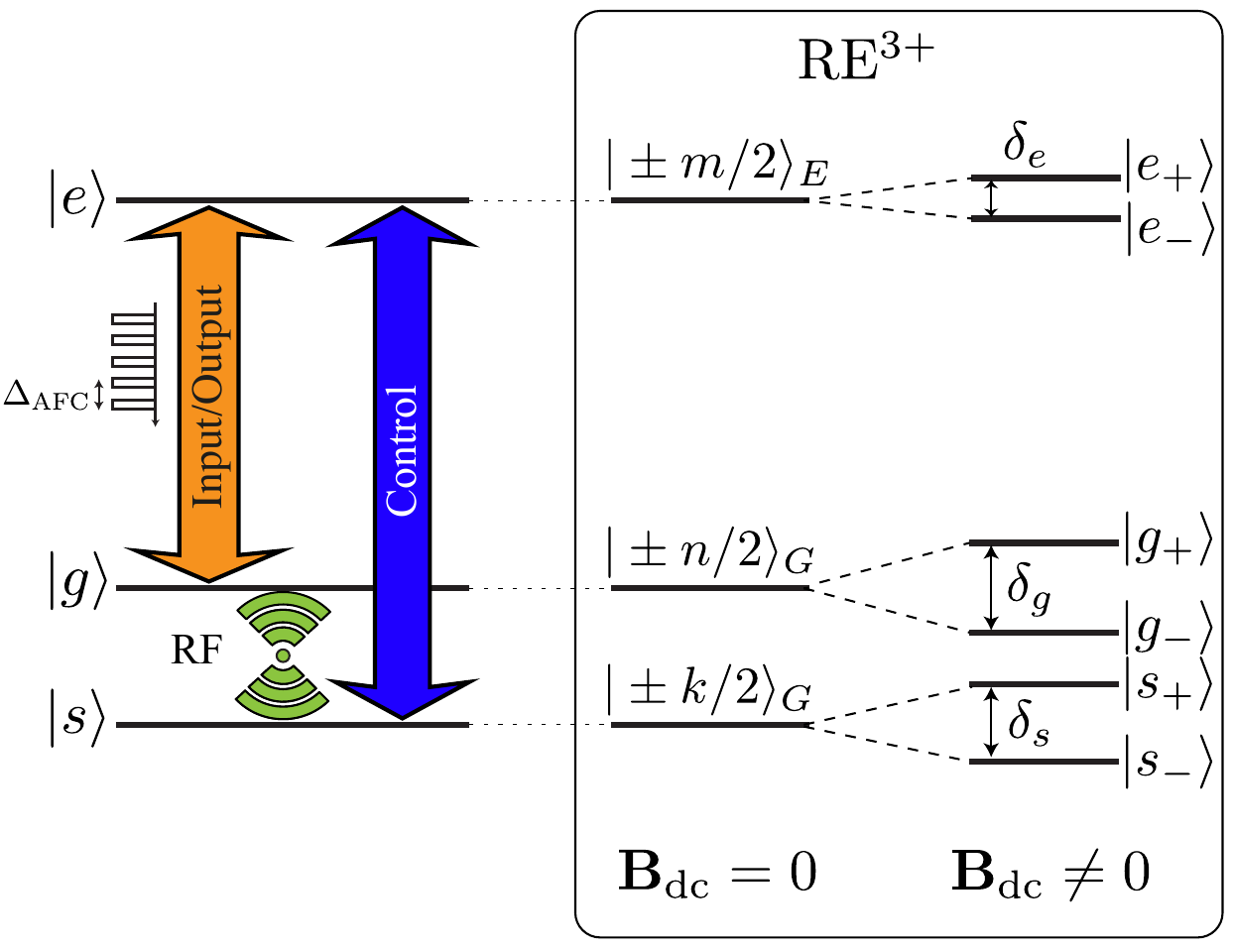}
\caption{The AFC protocol (left) and its application to nuclear spin levels of rare-earth ion doped crystals (right). Under the application of an external magnetic field $\mathbf{B}_{\rm dc}$, the different Zeeman levels split. }
\label{lvlscheme}
\end{center}
\end{figure}

Let us recall the principles of the spin-wave atomic frequency comb protocol. The protocol, as described for the first time in \cite{Afzelius09} and represented on the left part of figure \ref{lvlscheme}, relies on the shaping of the absorption profile of an ensemble of $N$ absorbers as a series of periodic teeth (period $\Delta_{\rm AFC}$) on a $\ket{g}\leftrightarrow\ket{e}$ transition. 

The absorption of an input photon by this structure then brings the ensemble in a so-called Dicke state:
\begin{align}
\ket{\psi}\propto\sum_{j=1}^Ne^{-i2\pi n_jt\Delta_{\rm AFC}}\ket{g_1,\dots,e_j,\dots,g_N},
\label{coh_opt}
\end{align}
where $n_j$ is the number of the tooth to which the j$^{th}$ atom belongs. According to this expression, the excitation will rephase after a time $\tau_{\rm AFC}=1/\Delta_{\rm AFC}$, leading to the re-emission of the input photon under the form of an echo. We will refer to this echo as the AFC echo. In forward emission, i.e. when the echo is emitted with the same $\mathbf{k}$ vector as the input pulse, the efficiency of retrieval can be written in a general form \cite{Jobez2016}:
\begin{equation}
\eta_{\rm AFC}=\eta_{\rm deph}\tilde{d}^2e^{-\tilde{d}}e^{-4/(T_2^{\rm opt}\Delta_{\rm AFC})}.
\label{eff_2lvl}
\end{equation}
In this expression, $\eta_{\rm deph}$ accounts for the remaining dephasing process due to the finite width and shape of the teeth of the comb, $\tilde{d}$ is the effective optical depth of the ensemble and $T_2^{\rm opt}$ is the optical coherence time. One can show that the optimal shape for the comb is square \cite{Bonarota:2010aa}, in which case $\tilde{d}=d/F$ ($d$ the optical depth, $F$ the comb finesse) and $\eta_{\rm deph}=\text{sinc}^2(\pi/F)$ \cite{Jobez2016}. The finesse that maximizes the efficiency is then $F=\pi/\text{atan}(2\pi/d)$ \cite{Bonarota:2010aa}.\\
In order to make this protocol on-demand, a third shelving level $\ket{s}$ can be used for the transfer of the optical coherence (\ref{coh_opt}) into a long-lived spin coherence. The efficiency of the full AFC spin-wave memory protocol is
\begin{align}
\eta_{\rm sw}=\eta_{\rm AFC}\eta_T^2e^{-2T_s/T_2^{\rm spin}},
\label{eff_3lvl}
\end{align}
where $\eta_T$ is the transfer efficiency from the optical to the spin transition, $T_s$ is the storage time, and $T_2^{\rm spin}$ is the spin coherence time \cite{Jobez2016}. Optical pumping techniques can be used to shape the frequency comb in the absorption profile \cite{Jobez2016}, and in this case a third ground state (that we will denote $\ket{\rm aux}$, not shown in figure \ref{lvlscheme}) has to be involved in order to store the removed population. These pumping techniques are impacted by the presence of a bias magnetic field, as we will see in section III. Additionally, spin manipulation may have to be performed between states $\ket{g}$ and $\ket{s}$ during the storage procedure to dynamically decouple the ions of interest from external fluctuations \cite{Holzapfel20}, or simply to rephase them efficiently due to the spin inhomogeneous linewidth \cite{Jobez15}. These manipulations are at the core of the study conducted in section \ref{sectionInversions} of the present article.
\subsection{Case of the REIDC}
Two main systems have demonstrated high performances regarding the AFC spin-wave protocol: europium (Eu$^{3+}$) and praseodymium (Pr$^{3+}$), by using the optical electronic $4f\leftrightarrow4f$ transition together with the nuclear quadrupole spin states. To be more explicit, the nuclear spin Hamiltonian of both electronic ground (noted $G$) and excited (noted $E$) states can be written in the form \cite{Teplov1968,Longdell2006,Lovric12,ZambriniCruzeiro2018a}:
\begin{subequations}
\begin{align}
H_{E}&=\mathbf{I}Q_{E}\mathbf{I}+\mathbf{B}(t)M_{E}\mathbf{I}\label{groundHamiltonian}\\
H_{G}&=\mathbf{I}Q_{G}\mathbf{I}+\mathbf{B}(t)M_{G}\mathbf{I}\label{groundHamiltonian},
\end{align}
\label{Hamiltonians}
\end{subequations}
\noindent where $\mathbf{I}$ is the nuclear spin vector, $Q_{X}$ are the effective pseudo quadrupole interaction tensors, $M_{X}$ the Zeeman tensors and $X$ stands for $E$ or $G$. $\mathbf{B}(t)$ is the applied magnetic field vector that can be decomposed into a constant component and an oscillatory component that drives the transitions between the states:
$\mathbf{B}(t)=\mathbf{B}_{\rm dc}+\mathbf{B}_{\rm ac}(t)$.
Then, the Hamiltonians (\ref{Hamiltonians}) can be decomposed as the sum of a time-independent and a time-dependent interaction Hamiltonian, respectively:
\begin{subequations}
\begin{align}
H_{X}=H_{X}^0+H_{X}^{\rm int}(t)
\end{align}
with
\begin{align}
&H_{X}^0=\mathbf{I}Q_{X}\mathbf{I}+\mathbf{B}_{\rm dc}M_{X}\mathbf{I}\label{hamilt_levels}\\
&H_{X}^{\rm int}(t)=\mathbf{B}_{\rm ac}(t)M_{X}\mathbf{I}.
\end{align}
\label{hamilt_gen}
\end{subequations}
In the case where $\mathbf{B}_{\rm dc}=0$, the two Hamiltonians (\ref{hamilt_levels}) give rise to $n=(2I+1)/2$ doubly degenerate states that can be used for the AFC protocol. This is shown in the central part of figure \ref{lvlscheme}, where two of these doubly degenerate states have been represented for the ground state ($\ket{\pm k/2}_G$ and $\ket{\pm n/2}_G$), and one for the excited state ($\ket{\pm m/2}_E$). For rare-earth-ions with nuclear spin $I$ above or equal to 5/2, the spin-wave AFC protocol can be implemented by using the three doubly degenerate ground states as the $\ket{g}$, $\ket{s}$ and the auxiliary level $\ket{\rm aux}$. Then, given the typical order of magnitude of the nuclear spin splittings, the spin manipulations can be performed by using radiofrequency (RF) pulses for inversion on the $\ket{g}\leftrightarrow\ket{s}$ transition.
Interestingly, the application of a magnetic field of the order of $\sim$ 10 mT to the ensemble leads to an increase in the coherence times \cite{Equall1994}, and allowed us to demonstrate spin-wave storage of classical optical pulses for about a second \cite{Holzapfel20}. The focus of the paper will be to understand how the spin and optical manipulation necessary for the memory protocols are affected by this external field, by only relying on the structure (\ref{Hamiltonians}) of the Hamiltonians, with a particular glimpse on the spin-wave AFC protocol. It has to be noted that most of the derivations that are performed throughout this article are not restricted to a specific ion in a given host crystal, but are valid for each system described by Hamiltonians of the form (\ref{Hamiltonians}), with $I\geqslant3/2$. Experimental verifications of the theoretical results have been conducted, with a setup that we will now describe.

\subsection{Experimental setup}
\begin{figure*}
\begin{center}
\includegraphics[width=17cm]{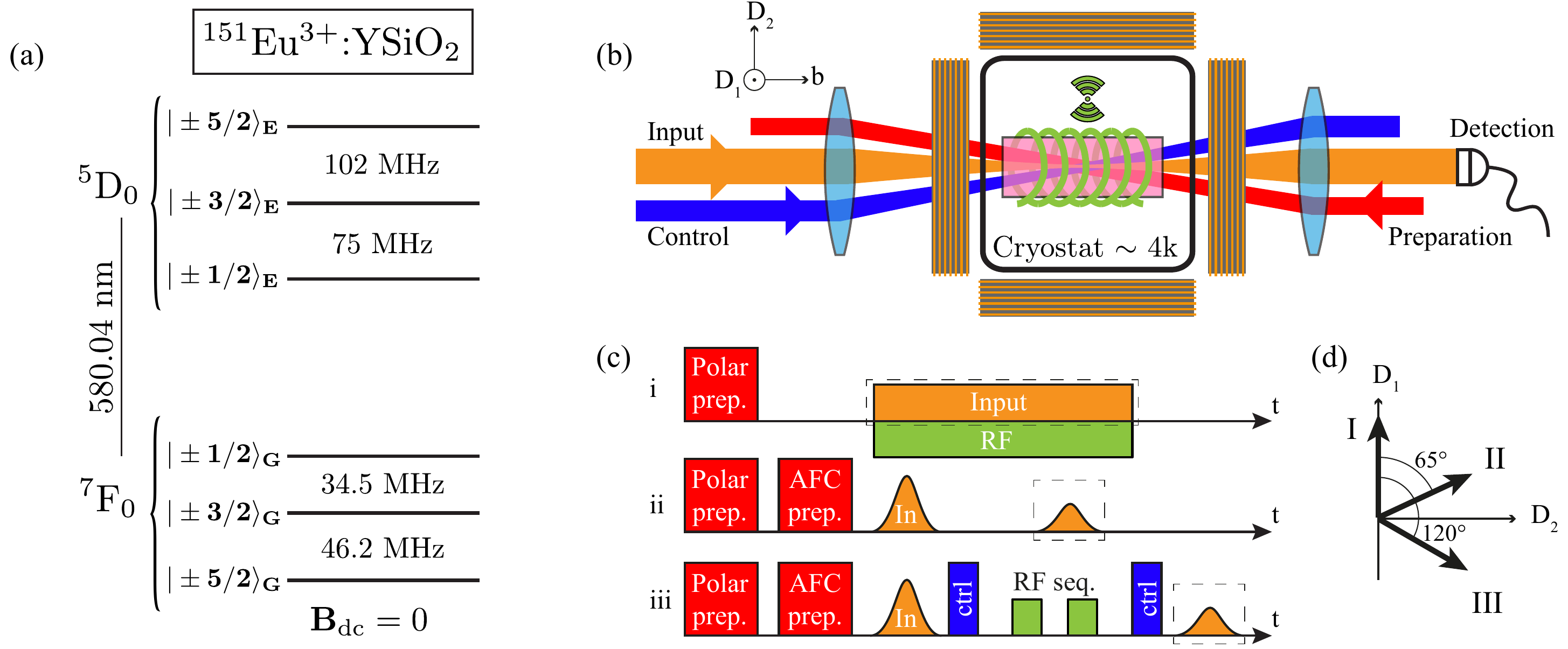}
\caption{Experimental system. (a) Energetic structure of ${}^{151}$Eu:Y$_2$SiO$_5$ at zero field. Application of a dc bias field will lift the degeneracy of the states, as shown on the right part of figure \ref{lvlscheme}. (b) Experimental apparatus. The b-cut ${}^{151}$Eu:Y$_2$SiO$_5$ crystal is placed in a cryostat at 4K, and optically addressed with three beams: preparation (red), input (orange) and control (blue). The RF coil (green), allows to apply the required RF fields. Three pairs of coils then allow to apply the required dc bias field to the crystal (the coils in the D$_1$ direction have not been represented here for clarity). The optical signal is detected with a variable gain photodiode in the input path. (c) Experimental sequences used throughout the article. The measured optical signal is surrounded with a dashed box. (i) Optically detected nuclear magnetic resonance: the spins are polarized in one of the ground states, and spin flops are performed with a constant amplitude RF field applied with the coil. Oscillations are detected with a simple absorption measurement, with the input beam. (ii) Simple AFC echo experiment: after preparation by optical pumping, an input pulse with a gaussian temporal shape is sent to the ensemble, and the intensity of the echo is measured.  (iii) Full spin-wave AFC protocol: same as in (ii) but with a transfer of the optical coherence in the spin domain, using the control fields (ctrl, in blue). RF pulses are applied during the spin-wave storage for rephasing of the inhomogeneous ensemble. (d) Definition of the three dc bias magnetic field directions I, II and III.}
\label{expscheme}
\end{center}
\end{figure*}

Our system is sketched in figure \ref{expscheme}. We use a 1000 ppm doped isotopically pure ${}^{151}$Eu:Y$_2$SiO$_5$ crystal, whose energy structure at zero field is shown in figure \ref{expscheme}(a). The host matrix Y$_2$SiO$_5$ is a biaxial crystal with polarization eigenaxes D$_1$, D$_2$, b \cite{Li1992}. The crystal is cut along these axes, and all optical beams are sent along the b axis, with their polarization along D$_1$ in order to maximize the absorption coefficient ($\alpha=2.6$ cm$^{-1}$). It is then placed on a home-built low-vibration mount in a closed-cycle helium cryostat at 4K. The 4f-4f $^7$F$_0$$\leftrightarrow$$^5$D$_0$ atomic transition is addressed with a laser at 580.04 nm.

The setup around the cryostat is shown in figure \ref{expscheme}(b). Thanks to acousto-optic modulators, the laser is split into three optical beams with arbitrary amplitude and phase control: the preparation, the input and the control. The preparation beam allows to shape the ensemble thanks to optical pumping techniques, and to prepare pits of absorption or complex structures like AFCs \cite{Jobez2016,Lauritzen12}. In particular, we use it prior to all our experiments to apply a class cleaning sequence, which allows to pump away all the atoms that are resonant on an unwanted transition. Therefore, after application of this sequence, individual $\ket{\pm k}_G\leftrightarrow\ket{\pm l}_E$ optical transitions can be addressed. Details of this procedure are given in \cite{Jobez2016,Lauritzen12} and shall not be discussed here. Then, the input beam is used either to probe the absorption of the ensemble or to send input pulses to be stored in it. Finally, the control beam allows to apply the optical transfer pulses that are required in the spin-wave AFC protocol. A six-turn coil wrapped around the crystal then allows to apply the radio-frequency ac field along the b axis for spin manipulation. In order to enhance the generated ac field, a resonator outside of the cryostat consisting in one capacitor in parallel and one in series is used \cite{Laplane2016a,Laplane2017,Holzapfel20}.\\
An external magnetic field can then be applied with three pairs of coils in Helmholtz configuration around the cryostat. With our setup, a maximum field amplitude of $\sim$~15 mT can be achieved. We label the direction of the field with the two angles $\theta_{\rm dc}$ and $\varphi_{\rm dc}$, such that $\mathbf{B}_{\rm dc}={B}_{\rm dc}[\cos(\theta_{\rm dc})\cos(\varphi_{\rm dc}),\cos(\theta_{\rm dc})\sin(\varphi_{\rm dc}),\sin(\theta_{\rm dc})]$ in the D$_1$, D$_2$, b basis. A particularity of the Y$_2$SiO$_5$ matrix subject to such a field is that there exists two magnetically inequivalent subsites for the ions, such that in general applying a dc bias field does involve four times as many levels as in the zero field case. However, due to the symmetries of the crystal, if $\mathbf{B}_{\rm dc}$ is applied along $b$ or in the (D$_1$,D$_2$) plane, the two sites behave equivalently and the system is simpler to handle \cite{ZambriniCruzeiro2018a}. In the rest of the manuscript, three particularly interesting field configurations within the (D$_1$,D$_2$) plane ($\theta_{\rm dc}=0$) are explored, and shown in figure \ref{expscheme}(d):
\begin{enumerate}
\item[-] \textbf{Direction I}, for which $\varphi_{\rm dc}=0$ (field aligned along D$_1$), where the effective gyromagnetic ratio of the ground state $\ket{\pm 1/2}_{G}$ is minimal ($\sim~4$~kHz/mT).
\item[-] \textbf{Direction II}, for which $\varphi_{\rm dc}=65^\circ$, where the effective gyromagnetic ratios of the ground states $\ket{\pm 1/2}_{G}$ and $\ket{\pm 3/2}_{G}$ are equal ($\sim~14$~kHz/mT).
\item[-] \textbf{Direction III}, for which $\varphi_{\rm dc}=120^\circ$, where the effective gyromagnetic ratio of the excited state $\ket{\pm 5/2}_{E}$ is minimal ($\sim~2.5$~kHz/mT).
\end{enumerate}

\section{Efficient population inversions in four-level systems}
\label{sectionInversions}

In order to perform spin echo sequences with high retrieval efficiency, high-quality rephasing sequences must be used. A common way to implement these sequences is to apply a series of inverting $\pi-$pulses. Among other parameters, the quality of inversion directly impacts the quality of the rephasing \cite{Zambrini16}. In this section, we are interested in understanding how one can perform efficient spin inversions between the $\ket{g}$ and $\ket{s}$ states. We will first answer this question for the zero field case, and generalize it for the presence of a field. In order to simplify the calculation, we will only consider the restriction of the spin Hamiltonian $H_G$ given in Eq. (\ref{groundHamiltonian}) to the subspace of dimension 4 spanned by $\{\ket{-k/2}_G,\ket{+k/2}_G,\ket{-n/2}_G,\ket{n/2}_G\}$, eigenvectors of $H^0_G$. In the following, and as shown in figure \ref{lvlscheme}, this basis will now be noted as  $\{\ket{s_-},\ket{s_+},\ket{g_-},\ket{g_+}\}$ for simplicity. We will note the restriction of all operators to this subspace via an exponent \{4\}. This restriction is equivalent to assuming that the population is constrained within the considered subspace, and that there is no leakage, by spontaneous emission, crosstalk, or any other process.

\subsection{Case $B_{\rm dc}$=0}
\label{zerofield}
\subsubsection{Hamiltonian}
When no external bias field is applied to the ions, the time-independent Hamiltonian $[H_{G}^0]^{\{4\}}$ reads:
\begin{align}
[H_{G}^0]^{\{4\}}=[\mathbf{I}Q_G \mathbf{I}]^{\{4\}}=-\frac{\hbar}{2}
\left[ 
\omega_0
\left(
\begin{array}{cc}
1\!\!1&0\\
0&-1\!\!1
\end{array}
\right)
\right],
\label{quadrupNofield}
\end{align}
where $\hbar\omega_0$ is simply the energy gap between the two doubly degenerate states $\ket{s_{\pm}}$ and $\ket{g_{\pm}}$ depicted in figure \ref{lvlscheme}, and $1\!\!1$ is the 2$\times$2 identity matrix. As $\mathbf{B}_{\rm dc}=0$, the magnetic field only consists of the driving radiofrequency field of the form
\begin{align}
\mathbf{B}(t)=\mathbf{B}_{\rm ac}(t)=B_{\rm ac}\cos(\omega_{\rm rot}t+\varphi)\mathbf{e_{\rm ac}},
\label{magfield_generic_ac}
\end{align}
where $B_{\rm ac}$, $\omega_{\rm rot}$, $\varphi$ and $\mathbf{e_{\rm ac}}$ are respectively the field amplitude, frequency, phase and direction ($||\mathbf{e_{\rm ac}}||=1$). One can show that in the same basis as in (\ref{quadrupNofield}), $[\mathbf{B}(t)M_{G}I]^{\{4\}}$ takes the form (see Appendix \ref{sectionUnitarity}):
\begin{align}
[\mathbf{B}_{\rm ac}(t)M_G \mathbf{I}]^{\{4\}}=-{B_{\rm ac}}\cos(\omega_{\rm rot}t+\varphi)
\left(
\begin{array}{cc}
G_{ss}&G_{sg}\\
G^{\dagger}_{sg}&G_{gg}
\end{array}
\right),
\label{matrixInteract}
\end{align}
where $G_{xy}=\frac{\mu_{xy}}{2}U_{xy}$. In this expression, $\mu_{sg}=2\sqrt{|\det(G_{sg})|}$ is the effective magnetic transition moment of the $s\leftrightarrow g$ transition and $U_{xy}$ are unitary matrices. Up to unitary transformations in the subsequent 2$\times$2 subspaces, one can choose $U_{sg}\in $ SU(2) (see Appendix \ref{sectionUnitarity}). For simplicity, we will now use the notation $U_{sg}:=U$ and $\mu_{sg}:=\mu$.
\subsubsection{Solution to the Schr\"odinger equation}
In order to understand the dynamics of the spin manipulation we simply have to solve the Schr\"odinger equation
\begin{align}
i\hbar\frac{d\ket{\psi_{\rm spin}(t)}}{dt}=H_G^{\{4\}}\ket{\psi_{\rm spin}(t)}
\label{SchroZeroField}
\end{align}
with $H_G^{\{4\}}=[\mathbf{I}Q_G\mathbf{I}]^{\{4\}}+[\mathbf{B}_{\rm ac}(t)M_G \mathbf{I}]^{\{4\}}$.
Then, in the rotating frame:
\begin{align}
\ket{\psi_{\rm spin}(t)}=
\left(
\begin{array}{cc}
e^{i\frac{\omega_{\rm rot}t}{2}}1\!\!1&0\\
0&e^{-i\frac{\omega_{\rm rot}t}{2}}1\!\!1
\end{array}
\right)\ket{\psi'_{\rm spin}(t)},
\end{align} 
and after doing the rotating wave approximation, (\ref{SchroZeroField}) becomes
\begin{align}
\frac{d\ket{\psi'_{\rm spin}(t)}}{dt}=\frac{i}{2}A\ket{\psi'_{\rm spin}(t)},
\label{eqpsi1}
\end{align}
where
\begin{align}
A=&
\Delta
\left(
\begin{array}{cc}
1\!\!1&0\\
0&-1\!\!1
\end{array}
\right)
+
\Omega_0
\left(
\begin{array}{cc}
0&e^{i\varphi}U\\
e^{-i\varphi}U^{\dagger}&0
\end{array}
\right),
\label{Amatrixzerofield}
\end{align}
$\Delta = \omega_{\rm rot}-\omega_0$ and $\Omega_0=\mu B_{\rm ac}/(2\hbar)$. In the case $\Delta=0$, the propagator associated with equation (\ref{eqpsi1}) is (see Appendix \ref{appendix_propag_low}):
\begin{align}
U_{0}(t)=\cos\left(\frac{\Omega_0t}{2}\right)1\!\!1+i~{\sin\left(\frac{\Omega_0t}{2}\right)}
\left(
\begin{array}{cc}
0&e^{i\varphi}U\\
e^{-i\varphi}U^\dagger &0
\end{array}
\right).
\label{spinPropNoB}
\end{align}
This propagator is similar to the case of a two-level drive, except each level is replaced by a doubly degenerate one. Consequently, perfect Rabi flops can be performed and ideal spin rephasing sequences can be applied. This means that the ground state depicted in figure \ref{lvlscheme} for $\mathbf{B}_{\rm dc}=0$ can indeed be treated as a two-level system regarding the population behavior during driving.\\
\subsubsection{Optical transitions}
Interestingly, a similar reasoning can be applied for the optical transition. Indeed, given the very different orders of magnitudes of the interactions, the nuclear spin contribution can be considered as a perturbation of the electronic part \cite{Bartholomew16}, which in turn allows to write the wavefunction of the ion as a tensor product between the electronic and the spin parts:
\begin{align}
\ket{\psi}=\ket{\psi_{\rm opt}}\otimes\ket{\psi_{\rm spin}}.
\end{align}
This allows to write the electric dipole interaction Hamiltonian for the optical fields following the notations in figure \ref{lvlscheme}, in the reduced basis  $\{\ket{s_-},\ket{s_+},\ket{g_-},\ket{g_+},\ket{e_-},\ket{e_+}\}$:
\begin{align}
H_{\rm opt}^{\{6\}}=d E_{\rm opt}\cos\left(\omega_{\rm rot}^{\rm opt}+\varphi^{\rm opt}\right)
\left(
\begin{array}{ccc}
0&0&G_{se}\\
0&0&G_{ge}\\
G_{se}^{\dagger}&G_{ge}^{\dagger}&0
\end{array}
\right),
\label{interact_hamilt_optic}
\end{align}
where $G_{ke}$ ($k$ standing for $g$ or $s$) are two-by-two matrices that we write of the form
$G_{ke}={b_{ke}}V_{ke}$. As for the magnetic transition, $b_{ke}^2=|\det G_{ke}|$ is the branching ratio of transition $k\leftrightarrow e$, and $V_{ke}$ are unitary matrices, of the form

\begin{subequations}
\begin{align}
V_{se}&=\frac{1}{b_{se}}
\left(
\begin{array}{cc}
\braket{s_-}{e_-}&\braket{s_-}{e_+}\\
\braket{s_+}{e_-}&\braket{s_+}{e_+}
\end{array}
\right)\\
V_{ge}&=\frac{1}{b_{ge}}
\left(
\begin{array}{cc}
\braket{g_-}{e_-}&\braket{g_-}{e_+}\\
\braket{g_+}{e_-}&\braket{g_+}{e_+}
\end{array}
\right).
\end{align}
\label{Unitary_optic}
\end{subequations}
The dynamics of the atom on the optical transition is then the one of a two-level system, as in (\ref{spinPropNoB}). Notice that in the present example, only one excited state and two ground states are considered, even if $n=(2I+1)/2$ nuclear states in each electronic state are present. In practice, all eigenstates are then vectors with $2n$ components.\\
\subsection{Case $B_{\rm dc}\neq0$} 
\subsubsection{Hamiltonian}
When subject to a constant dc magnetic field, the degeneracy of the doubly degenerate nuclear spin states is lifted. In the linear Zeeman regime, \textit{i.e.} when first order perturbation theory applies to the system (regime $\delta_x<<\omega_0$), the Hamiltonian (\ref{hamilt_levels}) in the basis $\{\ket{s_-},\ket{s_+},\ket{g_-},\ket{g_+}\}$ reads (see Appendix \ref{sectionUnitarity}):
\begin{align}
[H_{G}^0]^{\{4\}}=-\frac{\hbar}{2}\Bigg[ &\omega_0
\left(
\begin{array}{cc}
1\!\!1&0\\
0&-1\!\!1
\end{array}
\right)\nonumber\\
&+\delta_s
\left(
\begin{array}{cc}
\sigma_z&0\\
0&0
\end{array}
\right)
+\delta_g
\left(
\begin{array}{cc}
0&0\\
0&\sigma_z
\end{array}
\right)
\Bigg],
\label{NoInterHamiltBneq0}
\end{align}
where $1\!\!1$ is the two-by-two identity matrix and $\sigma_z$ is the usual Pauli matrix. Notice that we choose an increasing order in the eigenvalues, such that $\delta_s$ and $\delta_g$ are always positive. In the linear Zeeman regime, these splittings linearly depend on the field as $\delta_x~=~g_x~|B_{\rm dc}|$ (see Appendix \ref{sectionUnitarity}).\\
Then, following the same reasoning as in the previous paragraph, we have to solve the same equation (\ref{eqpsi1}) with 
\begin{align}
A=&
\Delta
\left(
\begin{array}{cc}
1\!\!1&0\\
0&-1\!\!1
\end{array}
\right)
+
\delta_s
\left(
\begin{array}{cc}
\sigma_z&0\\
0&0
\end{array}
\right)
+
\delta_g
\left(
\begin{array}{cc}
0&0\\
0&\sigma_z
\end{array}
\right)\nonumber\\
&+
\Omega_0
\left(
\begin{array}{cc}
0&e^{i\varphi}U\\
e^{-i\varphi}U^{\dagger}&0
\end{array}
\right)
\label{Amatrix}
\end{align}
and $\Omega_0=\mu B_{\rm ac}/(2\hbar)$. $U$ can be chosen in SU(2), similarly as in paragraph \ref{zerofield} (see Appendix \ref{sectionUnitarity}). It can then be written in the general form:
\begin{align}
U=\left(
\begin{array}{cc}
u_1&u_2\\
-u_2^*&u_1^*
\end{array}
\right),
\label{defin_unitary_matrices}
\end{align}
with $|u_1|^2+|u_2|^2=1$.\\
\subsubsection{Solution to the Schr\"odinger equation}
\begin{figure}[!h]
\begin{center}
\includegraphics[width=8.6cm]{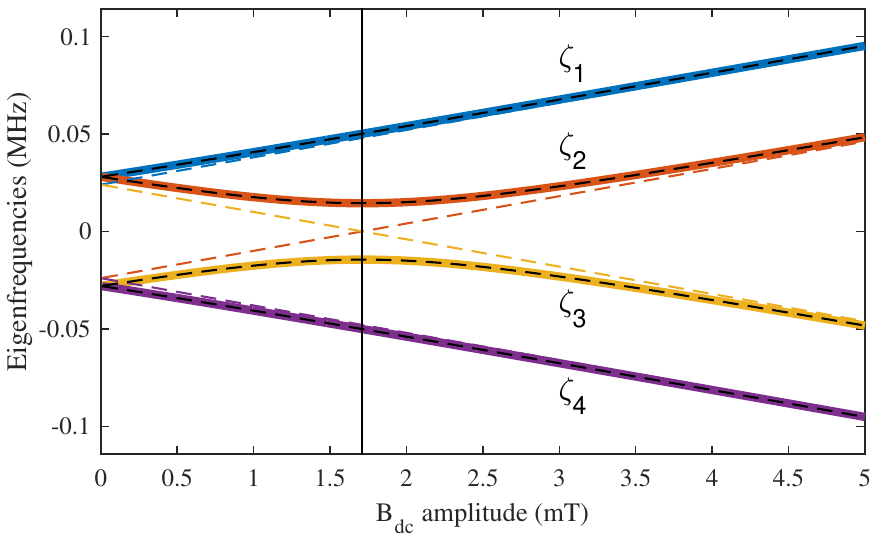}
\caption{Eigenvalues of the $A$ matrix given in Eq.(\ref{Amatrix}) as a function of the applied static magnetic field applied along direction II. A driving RF field, resonant with the spin transition and with 3 mT amplitude is applied along the $b$ axis. Solid lines are numerical solutions, black dashed lines are analytical approximations, given by Eq. (\ref{eigvalues}). Colored dashed lines are uncoupled eigenvalues ($|u_2|=0$, see Appendix \ref{sectionAmatrix})}
\label{anticross_eig}
\end{center}
\end{figure}

The solution of (\ref{eqpsi1}) is found by determining the four eigenvalues of $A$, which in the general case is a nontrivial problem. Fortunately, as shown in Appendix \ref{sectionAmatrix}, we can derive their approximate expression under the assumption  ${(g_s-g_g)^2}/|{g_sg_g}|~<<~1$, where $g_g$ (resp. $g_s$) is the effective gyromagnetic ratio of the $\ket{g}$ (resp. $\ket{s}$) state. This condition is fulfilled as soon as all gyromagnetic ratios are of the same order of magnitude, which is the case in most of the experimental situations for europium and praseodymium. In the case $\Delta=0$ a good approximation of the eigenvalues is given by:

\begin{subequations}
\begin{align}
\zeta_1&=\sqrt{\frac{\delta_g^2+\delta_s^2}{2}+\Omega_0^2+(\delta_g+\delta_s)\sqrt{\left(\frac{\delta_g-\delta_s}{2}\right)^2+|\Omega_1|^2}},\\
\zeta_2&=\sqrt{\frac{\delta_g^2+\delta_s^2}{2}+\Omega_0^2-(\delta_g+\delta_s)\sqrt{\left(\frac{\delta_g-\delta_s}{2}\right)^2+|\Omega_1|^2}},\\
\zeta_3&=-\zeta_2\\
\zeta_4&=-\zeta_1
\end{align}
\label{eigvalues}
\end{subequations}
where $\Omega_1=u_1\Omega_0$. The approximate values in the case $\Delta\neq0$ are given in Appendix \ref{sectionAmatrix} by Eqs.(\ref{eigenvalues_A}).\\
 If we apply this formula to our system with $\ket{s}=\ket{\pm3/2}_G$ and $\ket{g}=\ket{\pm1/2}_G$, we find the magnetic field dependency of the $\zeta_i$ shown in figure \ref{anticross_eig}. In the case chosen here, the $\mathbf{B}_{\rm dc}$ field is applied along direction II ($\theta_{\rm dc}=0$, $\varphi_{\rm dc}=65^\circ$), angle for which $g_g=g_s=~14$~kHz/mT, and the radio-frequency field is applied on resonance along the $b$ axis of the crystal (where $\mu/(2h)=10$~kHz/mT), with an amplitude of 3 mT, leading to a zero-field Rabi frequency of $\Omega_0=2\pi\times30$~kHz. For this field direction, the coupling terms are $|u_1|=0.856$ and $|u_2|=0.517$.

This behavior reveals an anti-crossing of amplitude $2|\Omega_2|$, where $\Omega_2=u_2\Omega_0$, at a magnetic field $B_{\rm cross}$ for which $|\Omega_1|~=~\sqrt{\delta_g\delta_s}$ (namely $B_{\rm cross}~=~B_{\rm ac}|u_1|\mu/\sqrt{g_gg_s}$). Intuitively, this new regime is entered when the system cannot be considered as a two-level system anymore: the Rabi frequency becomes comparable with the geometrical average of the splittings. The exact eigenvalues are represented in solid lines, while the approximations given by Eqs.(\ref{eigvalues}) are represented with dashed black lines, revealing the validity of the approximation. This figure also allows us to define three regions for the field amplitude that we will denote as:
\begin{enumerate}
\item[-] $B_{\rm dc}<<B_{\rm cross}$: weak field regime\\
\item[-] $B_{\rm dc}>>B_{\rm cross}$: strong field regime\\
\item[-] $B_{\rm dc}\sim B_{\rm cross}$: intermediate field regime.
\end{enumerate}
As we will see later, the behavior of the atoms in these three regions will be very different when attempting to perform population inversions.\\

Finally, the solution for Eq.(\ref{eqpsi1}) can be written in general as:
\begin{align}
\ket{\psi_{\rm spin}(t)}=&\alpha_{s_-}(t)\ket{s_-}+\alpha_{s_+}(t)\ket{s_+}\nonumber\\
&+\alpha_{g_-}(t)\ket{g_-}+\alpha_{g_+}(t)\ket{g_+},
\label{state_Rabi}
\end{align}
where
\begin{align}
\alpha_x(t)=\sum_{k=1}^4B_x^{(k)}e^{i\frac{\zeta_k}{2}t}.
\label{proba_amp_Rabi}
\end{align}
In this expression, the $B_x^{(k)}$ coefficients are uniquely determined by the initial conditions.

\subsubsection{ODNMR}
\begin{figure*}
\begin{center}
\subfigure[]{
\includegraphics[width=8.6cm]{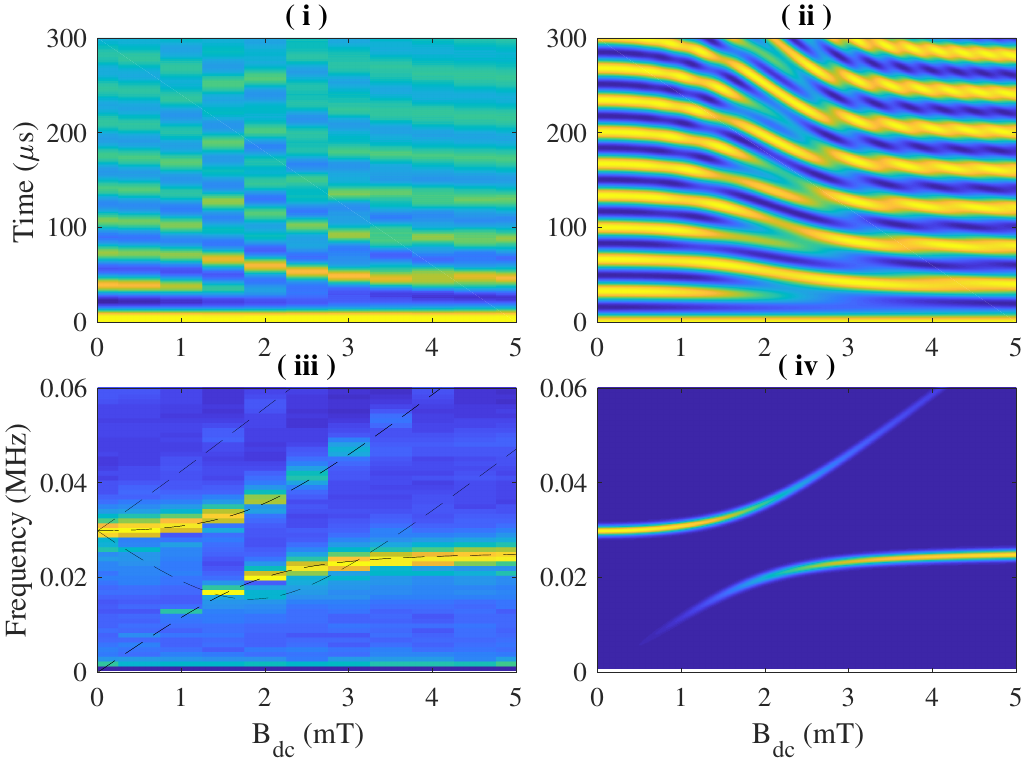}}
\subfigure[]{
\includegraphics[width=8.6cm]{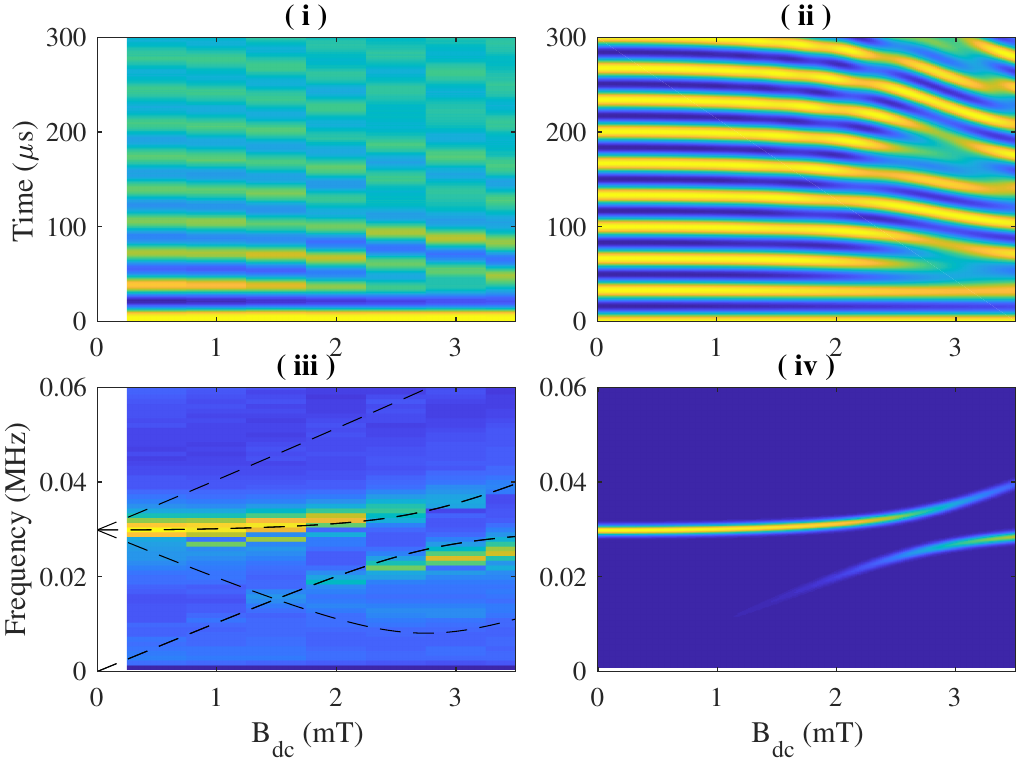}}
\caption{ODNMR traces for a dc bias magnetic field at (a) point II and (b) point III. In each case, (i) gives the experimental temporal traces as a function of the bias field and (iii) its Fourier transform, (ii) gives the expected analytical time trace as predicted by Eq.(\ref{intensODNMR}) and (iv) gives the Fourier transform of the theoretical curve. The experimental frequency trace is also overlapped with dashed lines, which indicate the expected positions of the frequency components. }
\label{anticross_exp}
\end{center}
\end{figure*}
In order to measure experimentally the previous parameters and validate the model, a simple method is to perform a drive of the spin population and follow the population dynamics by applying a constant amplitude RF field to the atoms. A common Optically Detected Nuclear Magnetic Resonance (ODNMR) technique for this purpose consists in driving the spin transition ($\ket{s}\leftrightarrow\ket{g}$ in figure \ref{lvlscheme}) while monitoring the absorption on an optical transition (e.g. $\ket{g}\leftrightarrow\ket{e}$ in figure \ref{lvlscheme}). Oscillations in the optical transmission directly give access to the spin population in the probed level, and thus allow to determine the population dynamics.\\
Referring to the previous analysis, the optical measurement will turn into measuring one population term $|\alpha_x(t)|^2$ from Eq.(\ref{proba_amp_Rabi}) or a sum of these populations when they cannot be discriminated. It directly becomes apparent that all the angular frequencies $\omega_{kl}=|\zeta_k-\zeta_l|$ could in principle appear in the observed oscillations. In the general case, this gives the possibility to observe up to 6 frequencies in the spin dynamics ($\omega_{12}$, $\omega_{13}$, $\omega_{14}$, $\omega_{23}$, $\omega_{24}$, $\omega_{34}$), which reduces to 4 in the case $\Delta=0$ ($\omega_{12}=\omega_{34}$ and $\omega_{13}=\omega_{24}$, see Eqs.(\ref{eigvalues})). 

We have performed the ODNMR protocol with our setup, of which the time sequence is shown in figure \ref{expscheme}(c), line (i). After the class-cleaning procedure, the population was polarized into the  $\ket{g}=\ket{\pm1/2}_G$ state with the preparation beam, and the coil generated a RF field at 34.54 MHz, resonant with the $\ket{g}\leftrightarrow\ket{s}=\ket{\pm3/2}_G$ transition ($\Delta=0$) at $B_{\rm dc}=0$. With our setup, a Rabi frequency at zero dc bias magnetic field of $\Omega_0=2\pi\times30$~kHz could be achieved. The optical input beam was then set to be resonant with the $\ket{g}\leftrightarrow\ket{e}=\ket{\pm5/2}_E$ transition. Notice that here, class cleaning is not performed at the Zeeman level, and thus the optical input beam simultaneously addresses four classes of atoms, each resonant with a different Zeeman transition. The two directions II and III were explored here. In particular, for direction III, the $|u_2|$ parameter in Eq.(\ref{defin_unitary_matrices}) is minimal, such that the avoided crossing has the smallest gap $2|\Omega_2|$ for a given ac field amplitude.\\
Figure \ref{anticross_exp}(a) and (b) parts (i) show the time traces of the ODNMR experiments as a function of the applied magnetic field $B_{\rm dc}$. The expected anti-crossing is made visible by performing the Fourier transform of the time traces, as shown in the plots (iii). Over these experimental curves, the four previously mentioned frequencies $\omega_{kl}$ are represented by black dashed lines, revealing a very good match with the measurement. However, it clearly appears that only two of them ($\omega_{12}$ and $\omega_{13}$) have a non-zero contribution. In order to understand this, we have used the analytical model developed in the previous paragraph (Eqs.(\ref{state_Rabi}-\ref{proba_amp_Rabi})) and made two hypotheses.
The first one is that, as no class cleaning is performed for the Zeeman levels, we suppose that the input beam simultaneously probes a class of atoms in $\ket{g_-}$ (noted $C_-$) and a class of atoms in $\ket{g_+}$ (noted $C_+$), such that the intensity of the transmitted input is proportional to  
\begin{align}
I(t)\propto\left|\alpha_{C_-,g_-}(t)\right|^2+\left|\alpha_{C_+,g_+}(t)\right|^2,
\label{intensODNMR}
\end{align}
 sum of the populations in the two states of the two classes, given by Eq.(\ref{proba_amp_Rabi}). The second hypothesis concerns the initial state of the spins: even if optical pumping is performed prior to the experiment in order to polarize the spins in $\ket{g}$, it does not discriminate between the Zeeman sublevels, such that we start with an initially mixed state between $\ket{g_-}$ and $\ket{g_+}$ for both classes $C_-$ and $C_+$:
 \begin{align}
 \rho_{C_-}(t=0)=\rho_{C_+}(t=0)=\frac12(\ket{g_-}\bra{g_-}+\ket{g_+}\bra{g_+}).
 \end{align}
 The corresponding theoretical time traces are shown in (ii) of figure \ref{anticross_exp} and show good agreement with the experimental ones (i). The theoretical Fourier components are displayed in (iv) and reveal indeed that the contributions for frequencies $\omega_{14}$ and $\omega_{23}$ of both classes of atoms annihilate:
 \begin{subequations}
\begin{align}
B_{C_-,g_-}^{(1)}\left(B_{C_-,g_-}^{(4)}\right)^*+B_{C_+,g_+}^{(1)}\left(B_{C_+,g_+}^{(4)}\right)^*&=0,\\
B_{C_-,g_-}^{(2)}\left(B_{C_-,g_-}^{(3)}\right)^*+B_{C_+,g_+}^{(2)}\left(B_{C_+,g_+}^{(3)}\right)^*&=0,
\end{align}
\end{subequations}
even if they are non-zero for each class. We note that this is true for $\Delta=0$ only.\\ It should finally be noted that in the experimental trace, the oscillations fade after $\sim100$~$\mu$s while they don't in the theoretical trace. Indeed, the spin manipulations are performed on an inhomogeneously broadened ensemble, such that each spin oscillates with its own Rabi frequency. Therefore, this decay depends both on the inhomogeneous broadening of the spin transition and on applied ac field (the spin homogeneous broadening contribution can be neglected at $\sim 100$~$\mu$s timescales). This decay gives the width of the frequency traces (iii) in figure \ref{anticross_exp}(a) and (b). However, in the theoretical model, the spin transition is considered to be perfectly homogeneously broadened and the theoretical frequency traces should consequently be infinitely thin. To make them visible, we have simply enlarged them with a gaussian profile with 700~Hz standard deviation width.\\
 
An important conclusion of this model is that if no care is taken about the magnetic field direction, amplitude and RF field detuning, up to 12 frequencies (6 per magnetic subsite) can simultaneously coexist in the population dynamics, clearly lowering the transfer efficiency of the RF pulses and altering their rephasing capabilities. In the two next paragraphs, we will address two possible strategies to circumvent this problem: one can either diminish the amplitude of the dc bias field in order to neglect its influence, or one can use complex adiabatic pulses that manage to drive the population efficiently.
 
\subsubsection{Efficient inversions in the weak field regime}
According to our previous considerations, if one applies a field $B_{\rm dc}<<B_{\rm cross}$, perfect inversions between the two states can be performed. This is what we observe experimentally from figure \ref{anticross_exp} with plots (i): at low fields, the population oscillates between $\ket{g}$ and $\ket{s}$ with a single frequency, and with a very good transfer efficiency. Let us prove mathematically this observation.
In the weak field regime, we can derive an analogous formula to Eq.(\ref{spinPropNoB}) for the propagator (see Appendix \ref{appendix_propag_low}):
\begin{align}
U_{\rm prop}(t)\simeq \cos\left(\frac{\epsilon|\Omega_1|t}{2}\right)U_0(t)+\sin\left(\frac{\epsilon|\Omega_1|t}{2}\right)U_{\rm pert}(t),
\label{propag_tot_maintext}
\end{align}
where $\epsilon=\frac{\delta_g+\delta_s}{2\Omega_0}<<1$, $U_0(t)$ is the propagator at zero field, given in Eq.(\ref{spinPropNoB}),

\begin{widetext}
\begin{align}
U_{\rm pert}(t)=&i\cos\left(\frac{\Omega_0t}{2}\right)
\left(
\begin{array}{cccc}
|u_1|&-u_2e^{i\phi_1}&0&0\\
-u_2^*e^{-i\phi_1}&-|u_1|&0&0\\
0&0&|u_1|&u_2e^{-i\phi_1}\\
0&0&u_2^*e^{i\phi_1}&-|u_1|
\end{array}
\right)
&+\sin\left(\frac{\Omega_0t}{2}\right)
\left(
\begin{array}{cccc}
0&0&-e^{i(\varphi+\phi_1)}&0\\
0&0&0&e^{i(\varphi-\phi_1)}\\
-e^{-i(\varphi+\phi_1)}&0&0&0\\
0&e^{-i(\varphi-\phi_1)}&0&0
\end{array}
\right),
\end{align}
\end{widetext}
and, $e^{i\phi_1}=u_1/|u_1|$.
By choosing a pulse duration of $\tau_{l}=(2l+1)\pi/\Omega_0$, the two-by-two diagonal blocks of $U_{\rm prop}$ vanish:
\begin{align}
U_{\rm prop}(\tau_l)\simeq& i(-1)^l \left(
\begin{array}{cc}
0&e^{i\varphi}\left(c_lU-is_lV\right)\\
e^{-i\varphi}\left(c_lU^\dagger-is_lV^\dagger\right) &0
\end{array}
\right),
\label{pipulseZeeman}
\end{align} 
with $c_l=\cos\left(\frac{\epsilon|\Omega_1|\tau_l}{2}\right)$ and $s_l=\sin\left(\frac{\epsilon|\Omega_1|\tau_l}{2}\right)$ and 
\begin{align}
V=\left(
\begin{array}{cc}
-e^{i\phi_1}&0\\
0&e^{-i\phi_1}
\end{array}
\right).
\end{align}
This form proves the aforementioned experimental observation, as no population remains in the initial state. It is also worth noticing that at first order the field does not induce an error on the quality of the inversion (the error is not equivalent to an imperfect rotation of angle $\pi+\delta\theta$) \cite{Zambrini16}.

\begin{figure*}
\begin{center}
\subfigure[Population in $\ket{-1/2}_G$]{
\includegraphics[width=8.6cm]{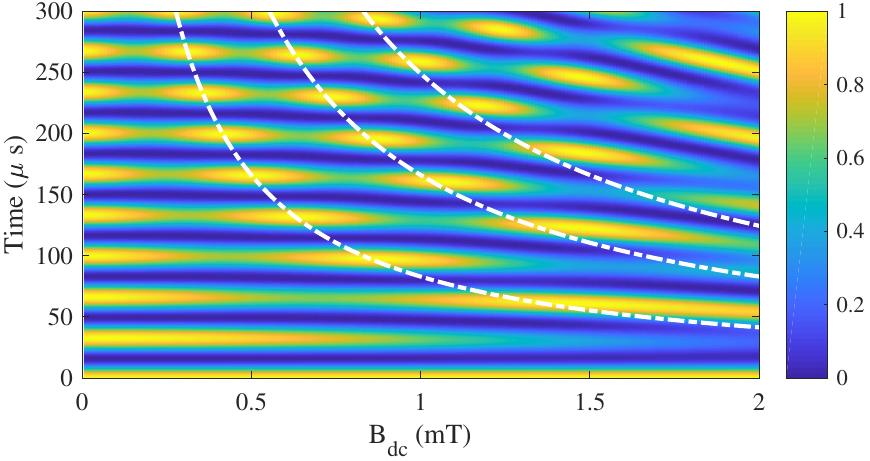}}
\subfigure[Population in $\ket{+1/2}_G$]{
\includegraphics[width=8.6cm]{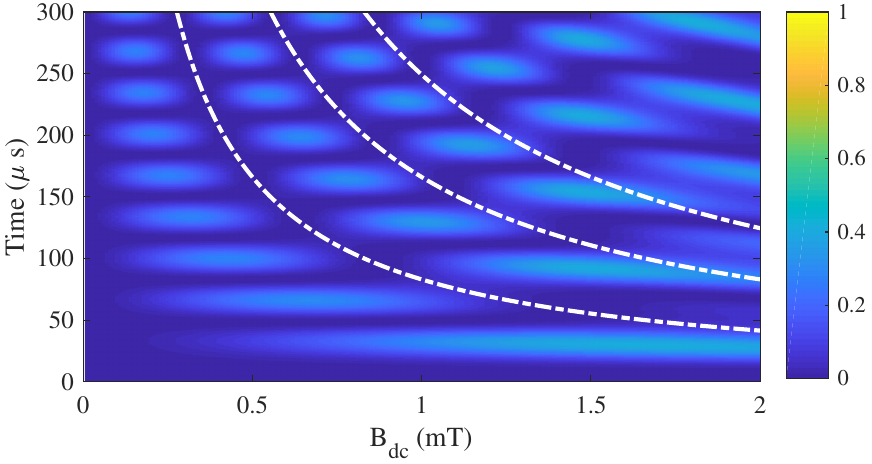}}
\subfigure[Population in $\ket{-3/2}_G$]{
\includegraphics[width=8.6cm]{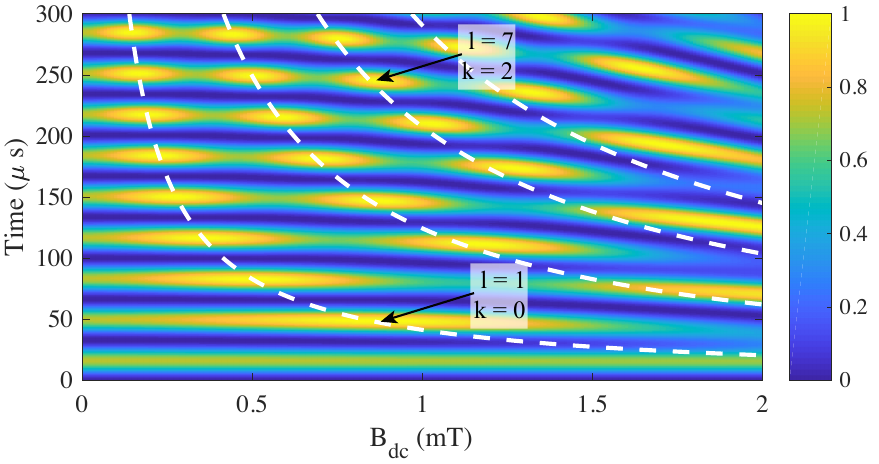}}
\subfigure[Population in $\ket{+3/2}_G$]{
\includegraphics[width=8.6cm]{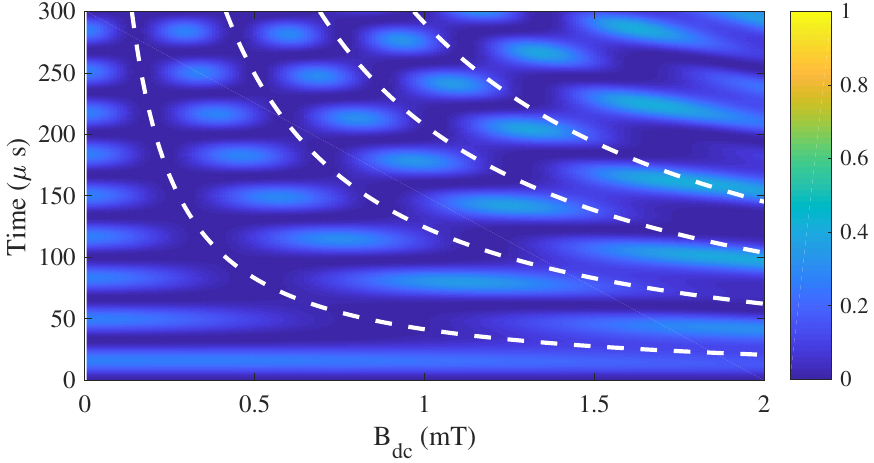}}
\caption{Numerical simulations of temporal spin population dynamics, for a continuously driven transition, as a function of applied dc bias magnetic field. Note that plot (ii) of figure \ref{anticross_exp}(a) is simply a sum of plots (a) and (b) here. The white dashed lines in (c) and (d) represent the condition at which the crosstalk between Zeeman sublevels vanishes as in equation (\ref{cond_for_max}) for $k$ between 0 and 3, highlighting regions where high quality inversions can be performed. The dashed-dotted lines in (a) and (b) represent twice the duration of the previous condition, highlighting regions where high-quality identity operation can be performed.}
\label{popdynZeem}
\end{center}
\end{figure*}
However, this expression also reveals that the inversion induces some intermixing in both $\ket{g_\pm}$ and $\ket{s_\pm}$ doublets. To be more explicit, up to a global phase and in the case $\varphi=0$ the top anti-diagonal term reads:
\begin{align}
c_lU-is_lV=&(\mathcal{R}(u_1)c_l-\sin\phi_1s_l)1\!\!1\nonumber\\&+i\mathcal{I}(u_2)c_l\sigma_x\nonumber\\
&+i\mathcal{R}(u_2)c_l\sigma_y\nonumber\\
&+i(\mathcal{I}(u_1)c_l+\cos\phi_1s_l)\sigma_z.
\end{align}
This matrix, also writable in the form 
\begin{align}
\exp\left(i\alpha\mathbf{n}\cdot\hat{\bm{\sigma}}/2\right)=\cos\frac{\alpha}{2}1\!\!1+i\mathbf{n}\cdot\hat{\bm{\sigma}}\sin\frac{\alpha}{2},
\end{align}
corresponds to a rotation on the Zeeman Bloch sphere with an angle $-\alpha$
\begin{subequations}
\begin{align}
\sin\frac{\alpha}2&=\sqrt{|u_2|^2c_l^2+(\mathcal{I}(u_1)c_l+\cos\phi_1s_l)^2}\\
\cos\frac{\alpha}2&=\mathcal{R}(u_1)c_l-\sin\phi_1s_l,
\end{align}
\end{subequations}
around the vector
\begin{align}
\mathbf{n}=\frac{1}{\sin\frac{\alpha}{2}}\left(
\begin{array}{c}
\mathcal{I}(u_2)c_l\\
\mathcal{R}(u_2)c_l\\
\mathcal{I}(u_1)c_l+\cos\phi_1s_l
\end{array}
\right).
\label{rotationNvector}
\end{align}

In the general case ($|u_2|\neq0$), this expression clearly indicates that a $(2l+1)\pi$-pulse also mixes the population of the Zeeman doublet: the rotation in the Bloch sphere does not occur around the $z$ axis. This is quite intuitive, as $|u_2|$ in the $U$ matrix is a coupling term between the two Zeeman states. To illustrate our mathematical finding, we plot in figure \ref{popdynZeem} the population evolution of an ion initially prepared in the $\ket{-1/2}_G$ state as a function of time and applied magnetic field, for $\ket{-1/2}_G$, $\ket{+1/2}_G$, $\ket{-3/2}_G$ and $\ket{+3/2}_G$. In this simulation, all parameters are identical to those of figure \ref{anticross_exp}(a), and note that plot \ref{anticross_exp}(a)(ii) is simply the sum of plots (a) and (b) in figure \ref{popdynZeem}. A striking fact that appears here is that opposed to figure \ref{anticross_exp}(a)(ii), the visibility of the Rabi oscillations is strongly affected by the field. Therefore, perfect transfer from $\ket{-1/2}_G$ to $\ket{-3/2}_G$ cannot be realized with a basic $\pi$ pulse at low field (see the first temporal population maximum in $\ket{-3/2}_G$ in plot (c), which is not 1).\\
The consequence of such a phenomenon is that the application of an even number of $\pi$ pulses together with free evolutions does not bring the spins back into their initial state in general, which is of crucial importance for rephasing sequences like spin-echo techniques \cite{Jobez15} or dynamical decoupling sequences \cite{Holzapfel20,Zambrini16}.
Fortunately, figure \ref{popdynZeem} also reveals that under specific conditions, near-perfect inversions without crosstalk can be performed in the weak field regime (regions simultaneously at maximal population for $\ket{-3/2}_G$ and minimal population for $\ket{+3/2}_G$). This is confirmed by equations (\ref{pipulseZeeman}) and (\ref{rotationNvector}): if $c_l=0$, the cross terms in the $c_lU-is_lV$ matrix vanish. This condition reads
\begin{align}
\frac{\epsilon|\Omega_1|\tau_l}{2}=(2k+1)\frac{\pi}{2},\qquad k\in\mathbb Z.
\label{cond_for_max}
\end{align}
and it is plotted in figure \ref{popdynZeem} (c) and (d) with dashed lines for $k$ between 0 and 3. In a way, the grid of peaks in the transfer efficiency is identified with an index $(l,k)$, where $l$ is the line number (from bottom to top), and $k$ is the column number (from left to right). Two points have for instance been highlighted in figure (c), one at $l=1$, $k=0$ and one at $l=7$, $k=2$. As expected, the model leads to a good prediction accuracy in the upper left corner of the plot, where  
\begin{align}
\epsilon=\frac{(2k+1)\pi}{|\Omega_1|\tau_l}=\frac{2k+1}{|u_1|(2l+1)}<<1,
\end{align}
where $k$ is small and $l$ is large. However, experimentally  there will obviously be a tradeoff between quality of the theoretical inversion and duration of the pulse, due to the finite coherence time. As an illustration, we see that our experimental curves have a contrast in the oscillations that vanish after $\sim$ 2 complete population inversions (see traces (i) of figure \ref{anticross_exp}), limiting $l$ in practice to 1 or 2. 
In summary, in order to perform inversions without crosstalk, we simply have to set the pulse duration to $\tau_{l}=(2l+1)\pi/\Omega_0$ and to adjust the dc bias field to 
\begin{align}
B_{\rm dc}=\frac{2(2k+1)\Omega_0}{(2l+1)(g_g+g_s)|u_1|}.
\label{BiasFieldCond}
\end{align}
For validation of our reasoning, we also plot in figure \ref{popdynZeem}(a) and (b) the expected position of the population after two optimal pulses, for $k=0$ to 2 with dash-dotted lines, and see that the population can be brought back with very high efficiency in $\ket{-1/2}_G$.

\subsubsection{Adiabatic transfer of population: transfers in all the regimes}
\label{subsubadiab}
\begin{figure}
\begin{center}
\subfigure[ ]{
\includegraphics[width=4cm]{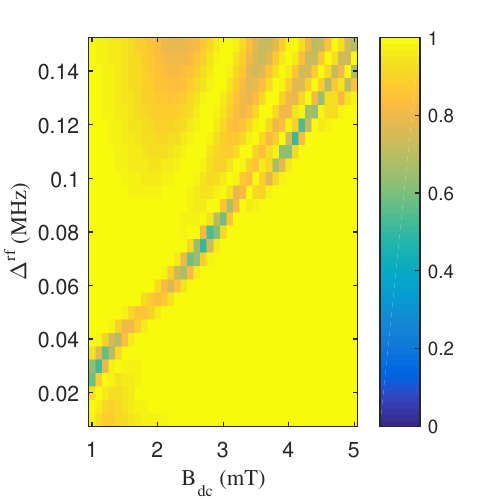}}
\subfigure[ ]{
\includegraphics[width=4cm]{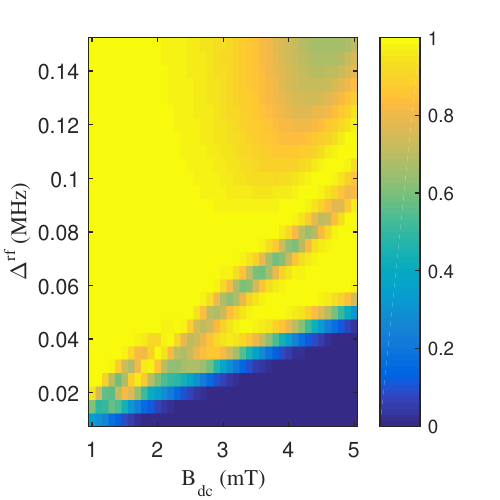}}
\caption{Numerical simulations of the population in $\ket{g}$ after adiabatic transfer from $\ket{s}$, with a secant hyperbolic profile (\ref{sech_profile}) where FWHM~=~120~\textmu s, as a function of $\Delta^{\rm rf}$ and the applied magnetic field along (a) direction II ($\varphi_{\rm dc}=65^\circ$) and (b) direction I ($\varphi_{\rm dc}=0^\circ$).}
\label{transfert_map}
\end{center}
\end{figure}

If we now release the constraint on the magnitude of the dc bias magnetic field, plots (ii) of figure \ref{anticross_exp} show that the pulse duration should be chosen very carefully in order to perform good quality inversions, due to the particularly messy transfer pattern around the anti-crossing point. Performing high quality inversions and rephasing \cite{Lauro:2011aa} is however usually possible by using adiabatic pulses, and we have numerically studied their application to an ion subject to a dc bias field. As before, we have focused on the simulation of our system, with the spin transition $\ket{g}=\ket{\pm1/2}_G\leftrightarrow\ket{s}=\ket{\pm3/2}_G$ at 34.54~MHz. The adiabatic pulses were chosen of the form \cite{Rippe2005,deSeze2005,Jobez2016}:
\begin{align}
B_{\rm ac}(t)=\text{sech}(\beta t)\cos\left[\omega_{\rm rot} t+\frac{\pi\Delta^{\rm rf}}{2\beta}\text{ln}\left(\cosh\left(\beta t\right)\right)\right],
\label{sech_profile}
\end{align}
where $\beta$ relates to the pulse full width at half maximum (FWHM) like $\beta=2\ \text{arch}(2)/{\rm FWHM}$, $\Delta^{\rm rf}$ is the pulse bandwidth (in Hz) and $\omega_{\rm rot}$ is the central angular frequency.\\
Figure \ref{transfert_map} shows the transfer map that can be achieved with such an adiabatic pulse of 120 $\mu$s FWHM and various chirps $\Delta^{\rm rf}$ as a function of the applied dc bias field. In the simulation, the population was initially set to an equal statistical mixture between $\ket{s_-}$ and $\ket{s_+}$, and we plot $|\alpha_{g_+}(\infty)|^2+|\alpha_{g_-}(\infty)|^2$, sum of the populations long after the pulse in $\ket{g_-}$ and $\ket{g_+}$.
A first striking fact is that imperfect inversion cannot simply be linked with the position of the previously described avoided crossing. Indeed, for a field oriented along direction II (fig.\ref{transfert_map}(a)), imperfect inversion does not only occur for a magnetic field amplitude of $B_{\rm dc}\sim1.7$ mT
 as predicted in fig.\ref{anticross_eig}, but instead forms a complex pattern, which depends on the total pulse chirp $\Delta^{\rm rf}$. We recall however that along direction II, $g_g=g_s$ such that perfect inversion can be performed if the field amplitude is high enough: both parallel transitions ($\ket{s_-}\leftrightarrow\ket{g_-}$ and $\ket{s_+}\leftrightarrow\ket{g_+}$) remain resonant, independently of the field intensity. However, if one goes out of this configuration, for instance with a dc bias field applied along direction I (along D$_1$, $\varphi_{\rm dc}=0^\circ$), one finds the pattern shown on fig.\ref{transfert_map}(b), for which transfer efficiency drops to 0 if the applied dc field is too high, while still following a complex shape for higher chirp. In this case the splitting is such that the applied field separates all the transitions too much, making it impossible to address all of them simultaneously with good efficiency.\\
It then clearly appears that in order to perform good quality population inversion with adiabatic pulses, one has to carefully adjust the different pulse parameters. Fortunately, it seems that in the regime of field explored here, there is always a set of parameters that allows to perform arbitrarily good population inversions.

\section{AFC protocol at $B_{\rm dc}\neq0$}

In this section we will focus on the influence of the field on the AFC protocol efficiency, by studying the preparation sequence and the spin-wave decay curves for a system subject to an external field.
\subsection{Comb shaping}
\label{subcombshape}
\begin{figure}[h]
	\includegraphics[width=0.4\textwidth]{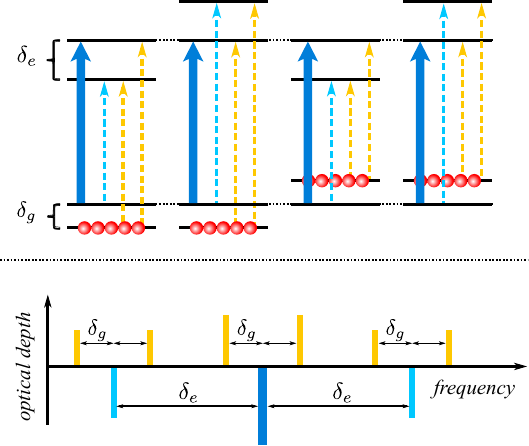}
	\caption{\label{FIG:hole-burning} 
		Hole burning in a four-level system. Pumping with a single frequency (wide dark blue arrow) creates a central transparency peak (large blue) as well as side transparency peaks at $\delta_e$ (thin blue, associated with blue dashed arrows). The pumped ions then lead to an increased absorption at frequencies $\delta_g$ around transparency peaks (orange, associated with orange dashed arrows).}
\end{figure}

If we remind what was briefly said in section \ref{sub_AFCprot}, the preparation of an AFC relies, in general, on two steps: first, the ensemble is polarized in the $\ket{g}$ state using optical pumping techniques, and second, spectral hole burning is used to burn the transparency regions of the AFC \cite{Jobez2016}. To this extent, ions at the corresponding frequencies are repeatedly excited to $\ket{e}$ until all of the unwanted population has relaxed to $\ket{\mathrm{aux}}$. If the degeneracy of $\ket{g}$ and $\ket{e}$ is lifted by an external magnetic field, the preparation of an AFC becomes more involved. To understand why, let us first consider the simple case of hole burning in an inhomogeneously broadened ensemble, as sketched in fig.\ref{FIG:hole-burning}. We will fist assume that the ensemble is only subject to an excited state splitting $\delta_e$. In this ensemble, any atom that is absorbing at frequency $f_0$ will also be absorbing either at frequency $f_0+{\delta}_e$ or at frequency $f_0-{\delta}_e$.
 Consequently, optical pumping at frequency $f_0$ will not only generate a transparency region at frequency $f_0$ (dark blue in figure \ref{FIG:hole-burning}), but also at frequencies $f_0\pm{\delta}_e$ (light blue)\cite{ZambriniCruzeiro2018a}. These additional transparency regions are usually referred to as `side holes'. If the bandwidth of the frequency comb is larger than ${\delta}_e$, then the side holes of the transparency regions of the AFC may overlap with the absorptive regions of the AFC. In this case, the optical depth of the comb is reduced and, consequently, its efficiency is decreased (see Eq.(\ref{eff_2lvl})). However if the excited state splitting is a multiple of the periodicity $\Delta_\mathrm{AFC}$ of the frequency comb:
\begin{equation}\label{EQU:sidehole2}
{\delta}_e=n\Delta_\mathrm{AFC}\quad\Longleftrightarrow\quad\Delta_\mathrm{AFC}=\frac{B_{\rm dc}{g_{e}}}{n} \quad n\in\mathbb{N},
\end{equation}
where $g_{e}$ is the effective gyromagnetic ratio of the excited state, then every side hole will coincide with a transparency region of the comb. In this case we expect no decrease of the memory efficiency compared to the degenerate case. 

Let us second consider the effect of a split ground state with splitting $\delta_g$. Population that is excited can then relax into the other ground state instead of $\ket{\mathrm{aux}}$ (see figure \ref{FIG:hole-burning}). Consequently, hole-burning at frequency $f_0$ will result in additional absorption at the frequencies $f_0\pm\delta_g$ (two orange central peaks in figure \ref{FIG:hole-burning}). One usually refers to these regions as `anti-holes'. Analogously to the scenario with a split excited state, one might expect that the efficiency of the comb is reduced if these anti-holes coincide with the transparent regions of the frequency comb. Consequently, an efficient comb can only be prepared if the anti-holes coincide with absorptive regions of the comb, leading to the condition

\begin{equation}\label{EQU:antihole2}
{\delta}_g=\left(n-\frac12\right)\Delta_\mathrm{AFC}\quad\Leftrightarrow\quad\Delta_\mathrm{AFC}=\frac{B_{\rm dc}{g_{g}}}{n-\frac12}\quad n\in\mathbb{N},
\end{equation}
where $g_g$ is the effective gyromagnetic ratio of the ground state.

Unlike the effect of side-holes, this does not necessarily pose a fundamental limitation. The comb preparation consists of many repeated cycles of excitation and relaxation, such that population that has relaxed back to $\ket{g}$ instead of $\ket{\mathrm{aux}}$ will be excited in the following cycles until it finally reaches $\ket{\mathrm{aux}}$. In this fashion the effect of anti-holes can be completely negated by repeating the preparation sequence sufficiently often, provided that $\ket{\mathrm{aux}}$ is long-lived enough that reflux from $\ket{\mathrm{aux}}$ to $\ket{g}$ can be neglected. In contrast to this, side-holes share the ground state with their respective central hole, such that transparency at the central hole position inevitably comes with increased transparency at the side-hole positions with a ratio between the two given purely by branching ratios, independently of the particulars of the preparation procedure.\\
Finally, if one considers a system with both a split ground- as well as excited state, additional absorption occurs at the frequencies $f_0\pm|\delta_g+\delta_e|$ and $f_0\pm|\delta_g-\delta_e|$ (four orange satellite peaks in figure \ref{FIG:hole-burning}). These anti-holes correspond to the transitions for which neither the ground state nor the excited state is shared with the central hole. Conditions similar to equation (\ref{EQU:antihole2}) may be formulated for these features, but for the same reason as mentioned previously, we do not expect to be fundamentally limited by these anti-holes.

\begin{figure}[ht!]
	\includegraphics[width=0.5\textwidth]{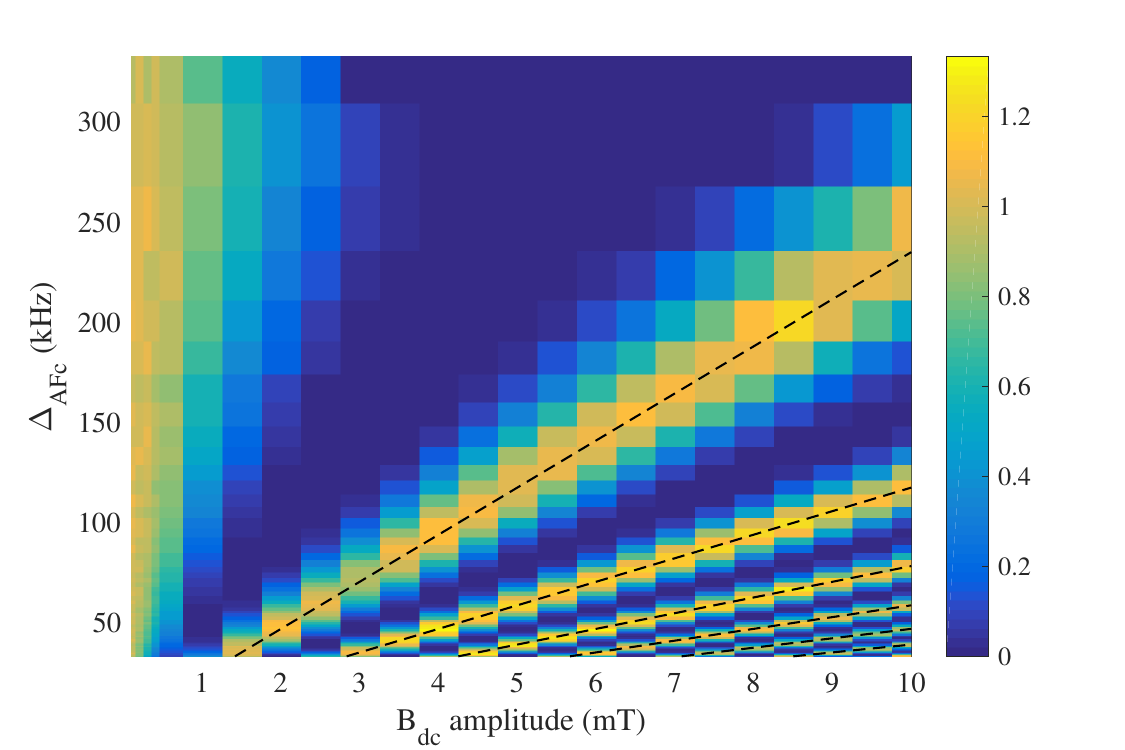}
	\caption{\label{FIG:D1} 
		Experimental efficiency ratio of AFCs prepared with and without an external field along direction I (D$_1$ direction, $\varphi_{\rm dc}=0^\circ$). We mark with dashed lines conditions (\ref{EQU:sidehole2}) for $n=1,...,6$ using an effective gyromagnetic ratio of $g_e=24$~kHz/mT, as determined in \cite{ZambriniCruzeiro2018a}.}
\end{figure}

In order to validate the previous prediction, we have performed simple optical AFC echo experiments under magnetic field with our system, by choosing $\ket{g}=\ket{\pm1/2}_G$ and $\ket{e}=\ket{\pm5/2}_E$. We have run the experiments with a field along direction I, where $g_e=24$~kHz/mT$>>g_g=4$~kHz/mT, allowing to test the validity of relation (\ref{EQU:sidehole2}), and with a field along direction III, where $g_g=12$~kHz/mT$>>g_e=2$~kHz/mT, allowing to test the validity of relation (\ref{EQU:antihole2}). The experimental time sequence that we use is shown in figure \ref{expscheme}(c) at line (ii), and simply consists in the preparation of an AFC and measurement of the two-level echo intensity for different values of the field applied during the whole sequence. To understand the relative effect of this field, we recorded the ratio between the AFC efficiency with and without it.\\
The data recorded along direction I are shown in figure \ref{FIG:D1} and reveal a very clear modulation of the efficiency: for some regions the efficiency is close to zero, while in other regions no decrease of the efficiency as compared to without bias field is observed. If we superimpose on the plot the condition that we have found for minimizing the disturbance of the preparation process (\ref{EQU:sidehole2}), we find that there is a very good agreement between the data and our expectation. Additionally, there is a region for low fields where the efficiency remains mainly unchanged. In this regime $\delta_e\ll\Delta_\mathrm{AFC}$, such that the side hole is sufficiently close to the central hole and its detrimental effect can be neglected. It has finally to be noted that as $g_g$ is very small for direction I, condition (\ref{EQU:antihole2}) is never met within the measured region. \\
\begin{figure}[ht!]
	\includegraphics[width=0.5\textwidth]{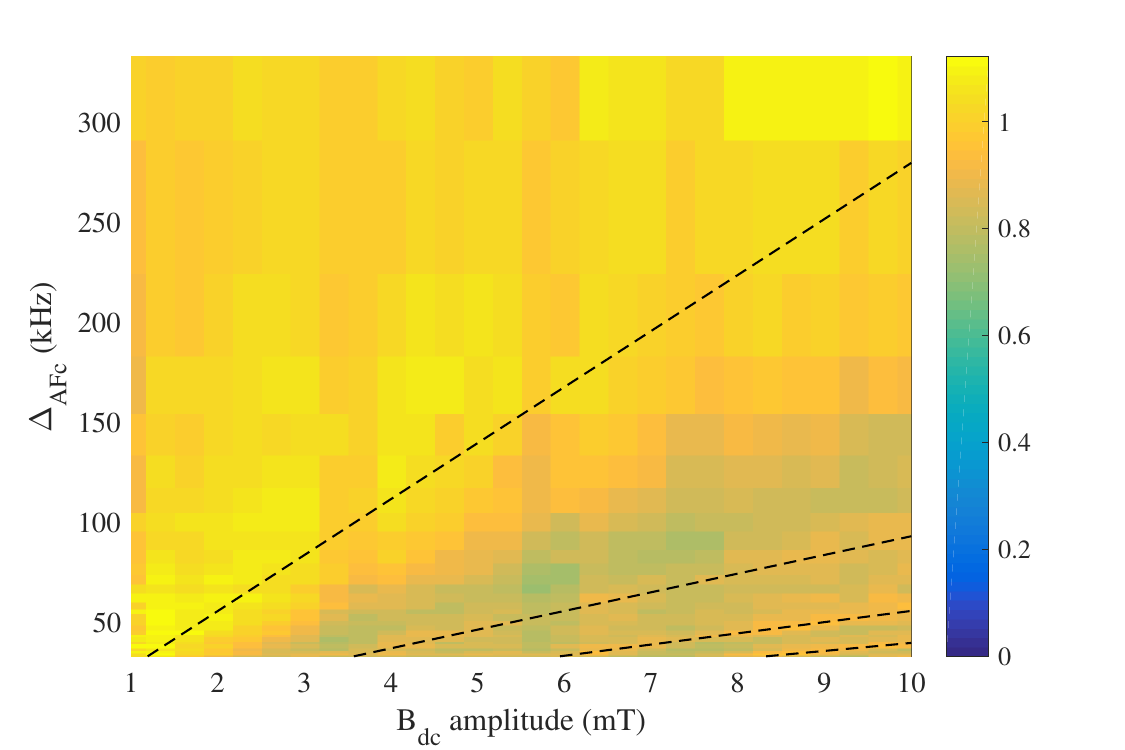}
	\caption{\label{FIG:120deg} 
Experimental efficiency ratio of AFCs prepared with and without an external field along direction III ($\varphi_{\rm dc}=120^\circ$). Compared to the previous case, $g_e<g_g$ such that the detrimental side holes are pushed away from the figure. A slight modulation is still visible, that we attribute to side anti-holes at $f_0\pm\delta_g$. The dashed lines indicate their expected positions, according to Eq. (\ref{EQU:antihole2}), for $n=1,...,4$ with $g_g=12$~kHz/mT, as determined in \cite{ZambriniCruzeiro2018a}.}
\end{figure}	

Along direction III, the measured ratios are shown in figure \ref{FIG:120deg} and clearly indicate a strongly reduced sensitivity of the efficiency on the magnetic field compared to point I, for all comb spacings. While the modulations that we expect from equation (\ref{EQU:sidehole2}) are outside of the observed parameter space, we notice another, much more faint modulation: comparing with the prediction from equation (\ref{EQU:antihole2}) seems to indicate that these modulations might be connected to the anti-hole at $f=f_0\pm\delta_g$, as depicted by the dashed lines. However, the strongly reduced strength of the modulation clearly corroborates our intuition that the side anti-holes do not play a role that is as detrimental as side-holes.\\
A simple way to link the two plots is that at point III, we remain in the zone $\delta_e<<\Delta_{\rm AFC}$ for all the fields under consideration, whereas the corresponding region was limited to the left part of the plot at point I. Following these observations, we note that efficient AFCs can be prepared in the presence of an external bias field even if their bandwidth is exceeding the Zeeman splitting that is induced by the field, provided that the periodicity of the comb is matched with the excited state splitting. The AFC preparation is therefore not fundamentally limiting the viability of AFC memories at low field in the REID systems that we consider.\\

\subsection{Spin-wave storage under weak magnetic field}
\begin{figure*}
\begin{center}
\subfigure[]{
\includegraphics[width=8.6cm]{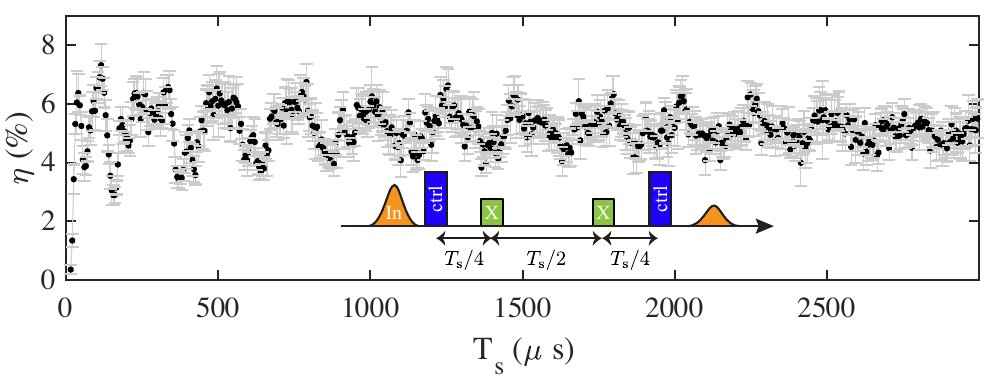}}
\subfigure[]{
\includegraphics[width=8.6cm]{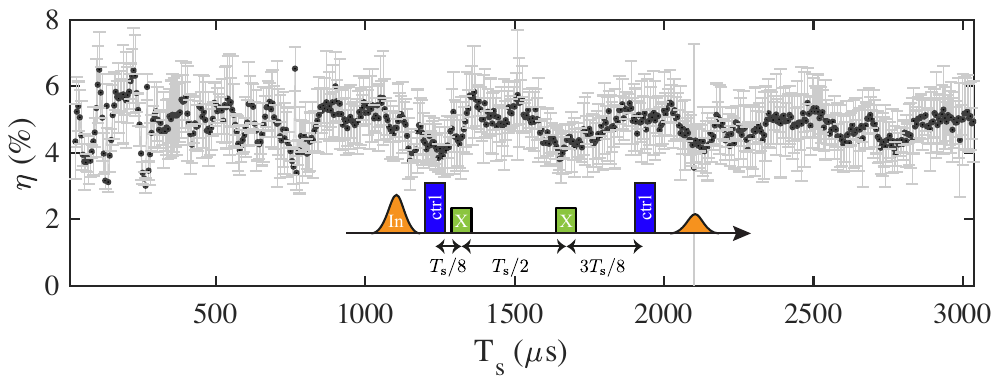}}
\subfigure[]{
\includegraphics[width=8.6cm]{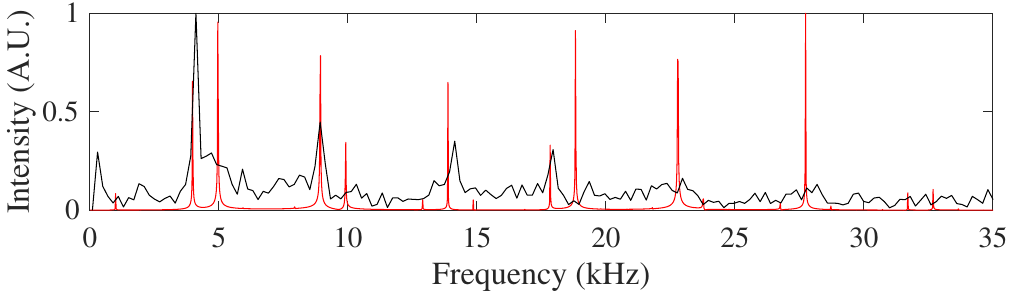}}
\subfigure[]{
\includegraphics[width=8.6cm]{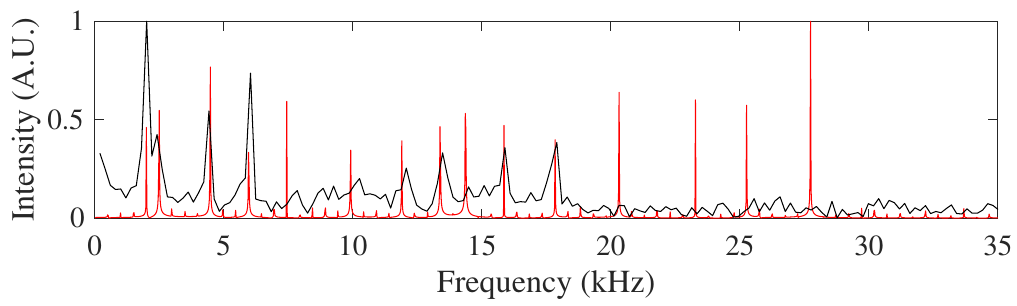}}
\caption{AFC spin wave echo modulations. (a) and (b): echo efficiency as a function of the spin-wave storage time $T_s$ under 1.4 mT bias magnetic field along direction I with a XX RF sequence that is (a) centered and (b) shifted by $T_s/8$. (c) and (d) Fourier transforms of (a) and (b) respectively. Numerical simulations are shown in red.}
\label{AFCSW}
\end{center}
\end{figure*}
Now that we have identified a favorable regime for the amplitude of the magnetic field regarding AFC preparation, let us focus on its influence on the full AFC spin-wave storage efficiency. We remind in figure \ref{expscheme}(c) plot (iii) the principle of the experiment: in addition to the two-level AFC protocol, optical control fields are used in order to store the coherence in a long-lived spin state. 
\subsubsection{Optical transfer pulses}
A question that arises is the influence of the field on the optical transfer pulse efficiency. Indeed, given the form of the optical interaction Hamiltonian (\ref{interact_hamilt_optic}), one should observe exactly the same phenomena as what was described in section \ref{sectionInversions} for the spin inversions. Fortunately, we can derive the same kind of condition as defined in previous section: for an optical drive on the $\ket{g}\leftrightarrow\ket{e}$ transition, the anti-crossing occurs at $\Omega^{\rm opt}_1=\sqrt{\delta_g\delta_e}$, where $\Omega^{\rm opt}_1=\braket{g_-}{e_-}dE^{\rm opt}/\hbar$. Notice the similarity with the writing $\Omega_1=u_1\mu B_{\rm ac}/\hbar$ for the spin transition, where $\mu=\mu_{sg}$ can be identified to $b_{ge}d$ and $u_1$ (first diagonal element of $U=U_{sg}$ in Eq.(\ref{defin_unitary_matrices})) can be identified to $\braket{g_-}{e_-}/b_{ge}$ (first diagonal element of $V_{ge}$ in Eq.(\ref{Unitary_optic})). Experimentally verifying this behavior in the optical domain is however more involved than in the spin domain as we have performed in section \ref{sectionInversions}, due to the much more limited coherence times: spin coherence times are for instance around three orders of magnitude larger than optical ones, whereas the Rabi frequencies are at best one order of magnitude better in the optical domain. We then only expect to see a decrease in the efficiency when approaching the avoided crossing point.\\
To be far from this region, one simply has to consider $\braket{g_-}{e_-}dE^{\rm opt}>>\sqrt{g_gg_e}B_{\rm dc}$, which is easily realized with optical field powers of $\sim100$~mW focused on $\sim 10$~\textmu m, leading for instance to optical Rabi frequencies of the order of $\sim 100$~kHz for europium. Then, one has to limit the $B_{\rm dc}$ field to get splittings smaller than this value, or use guided designs to confine the optical field more and push the Rabi frequency even higher \cite{Seri2018}. With this in mind, we investigated the shape of the decays of AFC spin-wave echoes.
\subsubsection{AFC spin-wave echoes}
Motivated by our recent study of dynamical decoupling under small magnetic field in the AFC spin-wave storage protocol \cite{Holzapfel20}, we have investigated the shape of the measured decay more deeply. For this purpose, we have used our europium sample in a slightly different experimental apparatus, and selected $\ket{\pm 5/2}_G$ as $\ket{g}$, $\ket{\pm 3/2}_G$ as $\ket{s}$ and $\ket{\pm 1/2}_E$ as $\ket{e}$ in order to implement the protocol (see figures \ref{lvlscheme} and \ref{expscheme} for notations). The spin manipulation now occurs at a RF frequency of 46.2 MHz, and is performed in the same way as depicted in previous part. Due to technical reasons, we had to use a different experimental apparatus as compared with previous section, and the field could only be applied along $D_1$ (direction I), and with a maximal amplitude of 1.4 mT. The AFC parameter for this experiment is $1/\Delta_{\rm AFC}=$~20~$\mu$s, and adiabatic optical transfer pulses were used \cite{Jobez2016}.\\
The time sequence that we have used in our protocol is shown in insets of figures \ref{AFCSW} (a) and (b). The overall efficiency (\ref{eff_3lvl}) of the protocol is $\eta_{\rm sw}\simeq5$~$\%$ in both cases, explained as follows. With our experimental parameters, the AFC echo efficiency (\ref{eff_2lvl}) is $\eta_{\rm AFC}=16$~\%. Then, in the AFC spin-wave memory efficiency formula (\ref{eff_3lvl}), two additional efficiency terms enter into play: the transfer efficiency $\eta_{T}$ and the spin dephasing term. In figure \ref{AFCSW} (a) and (b), we plot the efficiency of the spin-wave AFC echo as a function of the total storage time $T_s$ under a magnetic field of $B_{\rm dc}=1.4$~mT, and it clearly appears that the monotonically decreasing dephasing term does not play a role at the timescales we are considering here. Therefore, the remaining contribution should come from the transfer efficiency, which would be estimated to $\eta_{T}=\sqrt{\eta_{\rm sw}/\eta_{\rm AFC}}\simeq 56$~$\%$ according to Eq.(\ref{eff_3lvl}). However, the transfer pulses manage to reduce the amplitude of the AFC echo by 90~\%. Even if this value only gives an upper bound for its efficiency, the discrepancy between these two values is still unexplained and requires further investigations.\\
 Two scenarios are envisaged for the RF rephasing pulse configurations: in both cases a $XX$ RF sequence is used (two $\pi$ pulses with the same phase \cite{Jobez15}), but in (a) it is centered around $T_s/2$ whereas in (b) it is shifted by $T_s/8$. It clearly appears that the echo decays contain a complex oscillatory pattern, whose richness is made visible by their respective Fourier transforms in figures (c) and (d) in black. We propose here a simple numerical analysis to partially explain this shape, based on the study that we have performed in section \ref{sectionInversions}. We have considered a single initial spin in the $\ket{-5/2}_G$ initial state, and have applied to it a series of operators linked with Hamiltonian (\ref{interact_hamilt_optic}) for optical transfers, (\ref{pipulseZeeman}) for spin transfers, and free evolution operators in between, in order to simulate the whole spin-storage sequence. Namely, we simulate the output state as 
\begin{align}
\ket{\psi_{\rm out}}=&U_{\rm opt_2}^\pi U_{\rm spin}^{\rm free}(t_3)U_{\rm prop}(\tau_0)U_{\rm spin}^{\rm free}(t_2)U_{\rm prop}(\tau_0)U_{\rm spin}^{\rm free}(t_1)\nonumber\\
&\times U_{\rm opt_2}^\pi U_{\rm opt_1}^{\pi/2}\ket{\psi_{\rm in}}.
\end{align}
The initial coherence is simulated by a $\pi/2$ pulse for simplicity. Finally, the echo will be emitted with an amplitude proportional to the coherences weighted by their branching ratios.\\
We have plotted the numerical results in red for $2t_1=t_2=t_3=T_s/2$ in figure (c) and for $t_1=T_s/8$, $t_2=T_s/2$ and $t_3=3T_s/8$ in figure (d). We see with these plots that each experimentally observed peak can be associated with a numerically predicted one, and that the numerical simulation accounts for the difference between the two RF rephasing sequences. Given that the only difference between the two sequence is a phase accumulation difference, this clearly gives a solid hint that the oscillations originate from interferences between different quantum paths during spin storage.  Similar oscillations observed in stopped-light experiments in Pr:Y$_2$SiO$_5$ have also been interpreted as being due to nuclear Zeeman states split by small applied magnetic fields \cite{Heinze2011}.\\
The data presented in figure \ref{AFCSW} shows that the weights of the different frequency components are not well reproduced by the numerical model, even if actual branching ratios both in the optical and the spin domain have been taken into account. Also, unobserved additional high frequency components are predicted by the model.
These discrepancies could be due to specificities of the AFC protocol, and dephasing during optical evolutions. The model also considers that the whole ion coherence contributes to the emission of the echo, which is not strictly true for the AFC protocol, as the echo emission consists of the interference of the field emitted by many ions with different detunings: the single ion model is probably too simplistic here. Also, the optical control pulses that are actually used in the experimental setup are adiabatic pulses (see paragraph \ref{subsubadiab}), and the inversion dynamics is more complex than the one of a $\pi$ pulse as used in the numerical model. Despite these discrepancies, our simple toy model shows that the modulations in the echo have components that mostly originate from interferences between different Zeeman paths of a single ion.\\
These echo modulations originate from different phenomena as the ones witnessed with Kramers ions mostly originating from superhyperfine splitting \cite{Car2020}. Indeed, as europium does not possess an electronic spin, we expect this coupling to be of much smaller amplitude. However, oscillations due to superhyperfine interaction between Pr and Y ions have been observed in spin-echo measurements in Pr:Y$_2$SiO$_5$ \cite{Fraval2004b}. We note that Pr ions have larger nuclear magnetic moment than Eu ions in Y$_2$SiO$_5$. As the gyromagnetic ratio of Yttrium is $\gamma_Y=209$ Hz/G \cite{Fraval2004b}, if observable the associated oscillations should appear around 2 kHz in our case. In figure (c), we indeed see a weak amplitude peak at 2 kHz that is not explained by the previous numerical model, but in figure (d) other peaks linked with the numerical model hide its possible presence. To confirm this, we have also performed measurements with the same experimental apparatus as the one used in section \ref{subcombshape} for a dc bias magnetic field of 1 mT magnitude oriented along direction I and direction III and in each measurement a peak at $\sim 2$ kHz was indeed present. \\
As a remark, this study also highlights the fact that minimal crosstalk both in the spin and in the optical transitions have to be investigated together to find a fully-favorable configuration: condition (\ref{BiasFieldCond}) exhibited for the spin should hold simultaneously for the optical transition.

\section*{Conclusion}
We have investigated the effect of a dc bias magnetic field on different aspects of the manipulation of rare-earth ions, both in the optical and in the spin domain. After reminding the Hamiltonian of the considered category of ions, we have derived the solution of the Schr\"odinger equation for a four-level driven system, and have identified three field regimes (weak, strong, intermediate), that lead to different population dynamics. The theoretical model was compared to experimental realizations, and allowed to predict specific regimes in which perfect inversions could be performed. The effects of a dc bias magnetic field on the particular case of the spin-wave atomic frequency comb protocol were then tackled both for the preparation of the structure as well as for the explanation of complex oscillatory patterns in spin-wave echoes experiments.\\
The developed model should help to identify optimal field configuration in protocols involving optical or spin manipulation under magnetic field.
\section*{Acknowledgments}
We would like to thank Alexey Tiranov for fruitful discussions at the early stage of the developments presented here, as well as Claudio Barreiro for technical support. This work was financially supported by the European Union Horizon 2020 research and innovation program within the Flagship on Quantum Technologies through GA 820445 (QIA), by the Marie Sklodowska-Curie program through GA 675662 (QCALL) and by the Swiss FNS NCCR programme Quantum Science Technology (QSIT).
\appendix
\section{Form of the interaction matrix}
\label{sectionUnitarity}
In this appendix, we prove the form of the 2x2 block matrices that appear in the interaction matrix $\mathbf{B}M_X\mathbf{I}$ (see Eq. (\ref{matrixInteract})), once written in the eigenspace of $\mathbf{I}Q_X\mathbf{I}$.\\
Notice that the demonstration performed here does not require to assume the restriction to the space of dimension 4, but is valid for all half-integer spins. From now on, we also drop the index $X$ to simplify the writing.\\
Let us place ourselves in the eigenbase of $Q$, such that it can be written \cite{ZambriniCruzeiro2018a}:
\begin{align}
Q=
\left(
\begin{array}{ccc}
-E&0&0\\
0&E&0\\
0&0&D
\end{array}
\right).
\end{align}
Then, $\mathbf{I}Q\mathbf{I}$ takes the simple form:
\begin{align}
\mathbf{I}Q\mathbf{I}=-EI_x^2+EI_y^2+DI_z^2,
\label{appA_quadrup}
\end{align}
where the $I_i$ are the usual spin matrices, for a spin of value $I=n-1/2$, $n\in\mathbb N^*$.\\

\subsubsection{Rewriting of the spin matrices\\}
The first and most important step of this calculation is to group the spin $z-$projections two by two, thanks to the transformation
\begin{align}
F=
\left(
\begin{array}{ccccccccc}
1&&&&&&&&0\\
&0&&&&&&&1\\
&&1&&&&&0&\\
&&&\ddots&&&1&&\\
&&&&\ddots&&&&\\
&&&1&&&0&&\\
&&0&&&&&1&\\
0&1&&&&&&&0\\
\end{array}
\right).
\end{align}
Intuitively, this transformation exchanges the even indexes with respect to the end indexes (index 2 is exchanged with index $2n$, index $4$ is exchanged with index $2n-2$, etc.) while leaving the odd indexes untouched. In this way, the eigenvalues of the spin $z$-projection are grouped two by two, and we can find similarities of the spin matrices with the 1/2 spin Pauli matrices. To illustrate the effect of this transform, let us focus on the spin 3/2 case. Here, the transformation reads
\begin{align}
F=
\left(
\begin{array}{cccc}
1&0&0&0\\
0&0&0&1\\
0&0&1&0\\
0&1&0&0
\end{array}
\right),
\end{align}
such that the spin matrices are modified according to (remind that $F^{-1}=F$):
\begin{subequations}
\begin{align}
FI_xF&=\frac12
\left(
\begin{array}{cccc}
0&0&0&\sqrt{3}\\
0&0&\sqrt{3}&0\\
0&\sqrt{3}&0&2\\
\sqrt{3}&0&2&0
\end{array}
\right),\nonumber\\
&=\frac12
\left(
\begin{array}{cc}
0&\sqrt{3}\\
\sqrt{3}&2
\end{array}
\right)\otimes\sigma_x,
\end{align}

\begin{align}
FI_yF&=\frac12
\left(
\begin{array}{cccc}
0&0&0&-i\sqrt{3}\\
0&0&i\sqrt{3}&0\\
0&-i\sqrt{3}&0&2i\\
i\sqrt{3}&0&-2i&0
\end{array}
\right),\nonumber\\
&=\frac12
\left(
\begin{array}{cc}
0&\sqrt{3}\\
\sqrt{3}&-2
\end{array}
\right)\otimes\sigma_y,
\end{align}

\begin{align}
FI_zF&=\frac12
\left(
\begin{array}{cccc}
3&0&0&0\\
0&-3&0&0\\
0&0&-1&0\\
0&0&0&1
\end{array}
\right),\nonumber\\
&=\frac12
\left(
\begin{array}{cc}
3&0\\
0&-1
\end{array}
\right)\otimes\sigma_z.
\end{align}
\end{subequations}
This clearly shows that after the transform $F$, each spin matrix is composed of blocks of 2x2 matrices that are simply multiples of Pauli matrices.
This writing can then easily be generalized to arbitrary spin $I=n-1/2$:
\begin{subequations}
\begin{align}
FI_xF&=\frac12
\left(
\begin{array}{ccccc}
&&&&a_{2n-1}\\
&0&&a_{2n-3}&a_{2n-2}\\
&&\reflectbox{$\ddots$}&\reflectbox{$\ddots$}&\\
&a_3&a_4&&0\\
a_1&a_2&&&\\
\end{array}
\right)\otimes\sigma_x\\
&:=A_x\otimes\sigma_x,\nonumber
\end{align}

\begin{align}
FI_yF&=\frac12
\left(
\begin{array}{ccccc}
&&&&a_{2n-1}\\
&0&&a_{2n-3}&-a_{2n-2}\\
&&\reflectbox{$\ddots$}&\reflectbox{$\ddots$}&\\
&a_3&-a_4&&0\\
a_1&-a_2&&&\\
\end{array}
\right)\otimes\sigma_y\\
&:=A_y\otimes\sigma_y\nonumber,
\end{align}

\begin{align}
FI_zF&=\frac12
\left(
\begin{array}{ccccccc}
c_1&&&&\\
&c_3&&&0\\
&&\ddots&&\\
&0&&c_{2n-3}&\\
&&&&c_{2n-1}\\
\end{array}
\right)\otimes\sigma_z\\
&:=A_z\otimes\sigma_z,\nonumber
\end{align}
\label{appA_spinop}
\end{subequations}
where
\begin{subequations}
\begin{align}
a_k&=\sqrt{k(2n-k)}=a_{2n-k}\\
c_k&=2(n-k)+1,
\end{align}
\end{subequations}
and where the $\sigma_i$ are the Pauli matrices.\\

\subsubsection{Diagonalization of the quadrupole Hamiltonian\\}
Once written under the previous form, the squared spin matrices that appear in Eq.(\ref{appA_quadrup}) simply read
\begin{subequations}
\begin{align}
FI_x^2F&=A_x^2\otimes1\!\!1\\
FI_y^2F&=A_y^2\otimes1\!\!1\\
FI_z^2F&=A_z^2\otimes1\!\!1,
\end{align}
\end{subequations}
which gives the simple expression for the quadrupole component in the new basis:
\begin{equation}
F\mathbf{I}Q\mathbf{I}F=\left(-EA_x^2+EA_y^2+DA_z^2\right)\otimes1\!\!1.
\end{equation}
Then, in order to find the eigenvalues of this Hamiltonian, one just has to find the eigenvalues of the real symmetric matrix $-EA_x^2+EA_y^2+DA_z^2$. In other words, the matrix that will diagonalize $F{\mathbf{I}Q\mathbf{I}}F$ will only act on the left part of the tensor product, and can be written of the form $P\otimes1\!\!1$, where $P$ is a real orthogonal matrix of size $n$. This means that 
\begin{equation}
P^{-1}\left(-EA_x^2+EA_y^2+DA_z^2\right)P\otimes1\!\!1
\end{equation}
is diagonal, with $n$ doubly degenerate eigenvalues. This is the result that we expect, as plotted in figure \ref{lvlscheme}(right) and \ref{expscheme}(a) for the case of europium.\\

\subsubsection{Rewriting of the interaction Hamiltonian in the case $B_{\rm dc}=0$\\}
Let us now see the effect of these transforms on the interaction Hamiltonian $\mathbf{B_{\rm ac}}(t)M\mathbf{I}$. According to the general writing of $\mathbf{B_{\rm ac}}(t)$ of Eq.(\ref{magfield_generic_ac}), the interaction matrix form will be determined by the direction of the magnetic field $\mathbf{e}_{\rm ac}$. Mathematically, it can always be written as
\begin{align}
\mathbf{e}_{\rm ac}M\mathbf{I}=\alpha_x^{\rm ac}I_x+\alpha_y^{\rm ac}I_y+\alpha_z^{\rm ac}I_z,
\label{appA_interform_origin}
\end{align}
where $\alpha_i$ are real coefficients. Then, the transformation $F$ will simply affect the spin operators according to the (\ref{appA_spinop}) equations:
\begin{align}
F{\mathbf{e}_{\rm ac}M\mathbf{I}}F=\alpha_x^{\rm ac}A_x\otimes\sigma_x+\alpha_y^{\rm ac}A_y\otimes\sigma_y+\alpha_z^{\rm ac}A_z\otimes\sigma_z.
\end{align}
Then, the transformation $P\otimes1\!\!1$ which diagonalizes $F{\mathbf{I}Q\mathbf{I}}F$ acts on $F{\mathbf{e}_{\rm ac}M\mathbf{I}}F$ according to:
\begin{align}
&\left(P^{-1}\otimes1\!\!1\right)F{\mathbf{e}_{\rm ac}M\mathbf{I}}F\left(P\otimes1\!\!1\right)\nonumber\\
&=\alpha_x^{\rm ac}P^{-1}A_xP\otimes\sigma_x+\alpha_y^{\rm ac}P^{-1}A_yP\otimes\sigma_y+\alpha_z^{\rm ac}P^{-1}A_zP\otimes\sigma_z\nonumber\\
&:=\mathcal{A}_x^{\rm ac}\otimes\sigma_x+\mathcal{A}_y^{\rm ac}\otimes\sigma_y+\mathcal{A}_z^{\rm ac}\otimes\sigma_z.
\label{appA_interform_changed}
\end{align}
Formally speaking, the generic form of each 2x2 submatrix in Eq.(\ref{appA_interform_changed}) is then simply
\begin{subequations}
\begin{align}
G_{kl}^{\rm ac}&=\left(\mathcal{A}_x^{\rm ac}\right)_{kl}\sigma_x+\left(\mathcal{A}_y^{\rm ac}\right)_{kl}\sigma_y+\left(\mathcal{A}_z^{\rm ac}\right)_{kl}\sigma_z\label{appA_generform_inter}\\
&=\frac{\mu_{kl}}{2}U_{kl}^{\rm ac},
\end{align}
\end{subequations}
with 
\begin{align}
\mu_{kl}&=2\sqrt{-\det(G_{kl}^{\rm ac})}\nonumber\\&=2\sqrt{\left(\mathcal{A}^{\rm ac}_x\right)_{kl}^2+\left(\mathcal{A}^{\rm ac}_y\right)_{kl}^2+\left(\mathcal{A}^{\rm ac}_z\right)_{kl}^2}
\end{align}
the effective magnetic moment of the transition,
\begin{align}
U_{kl}^{\rm ac}&=\left(\mathcal{C}^{\rm ac}_x\right)_{kl}\sigma_x+\left(\mathcal{C}^{\rm ac}_y\right)_{kl}\sigma_y+\left(\mathcal{C}^{\rm ac}_z\right)_{kl}\sigma_z,\label{appA_interform_changed2}
\end{align} 
and $\left(\mathcal{C}^{\rm ac}_i\right)_{kl}=2\left(\mathcal{A}^{\rm ac}_i\right)_{kl}/\mu_{kl}$ such that $U_{kl}$ is a unitary matrix with determinant -1.\\
It is worth noticing that one can also perform arbitrary unitary transforms $\mathfrak{U}_i$ in each of the 2x2 subspaces, resulting in the redefinition of the $U_{kl}^{\rm ac}$ matrices:
\begin{align}
\left(U_{kl}^{\rm ac}\right)'=\mathfrak{U}^\dagger_kU_{kl}^{\rm ac}\mathfrak{U}_l.
\label{appA_transfo_inter}
\end{align} 
In particular, in the body of the article, we chose $U_{sg}$ ($k=1$ and $l=2$ in the spin 3/2 case) such that $U_{sg}~\in ~SU(2)$. This can be done from Eq.(\ref{appA_interform_changed2}) and (\ref{appA_transfo_inter}) by using $\mathfrak{U}_k=1\!\!1$ and $\mathfrak{U}_l=\sigma_z$ such that
\begin{align}
\left(U_{kl}^{\rm ac}\right)'=-i\left(\mathcal{C}_x^{\rm ac}\right)_{kl}\sigma_y+i\left(\mathcal{C}_y^{\rm ac}\right)_{kl}\sigma_x+\left(\mathcal{C}_z^{\rm ac}\right)_{kl}1\!\!1,
\end{align}
which clearly has a determinant +1.
\subsubsection{Influence of a magnetic field\\}
The application of an external magnetic field $\mathbf{B}_{\rm dc}~=~{B}_{\rm dc}\mathbf{e}_{\rm dc}$ on our system modifies the Hamiltonian according to Eq.(\ref{hamilt_levels}):
\begin{align}
H^0=\mathbf{I}Q\mathbf{I}+{B}_{\rm dc}\mathbf{e}_{\rm dc}M\mathbf{I}.
\label{appA_hamilt_BDC}
\end{align}
If we suppose that the applied magnetic field is sufficiently small, i.e. that the Zeeman splittings are all small compared to the quadrupolar splittings, then one can perform first order perturbation theory to estimate the effect of the field on the eigenlevels as well as eigenstates. To this extent, we simply have to diagonalize the 2x2 submatrices of $\mathbf{B}_{\rm dc}M\mathbf{I}$ corresponding to the eigenspaces of $\mathbf{I}Q\mathbf{I}$, found in the previous paragraphs. We recall that the diagonal matrix  that we found is $\left(P^{-1}\otimes1\!\!1\right)F\mathbf{I}Q\mathbf{I}F\left(P\otimes1\!\!1\right)$.\\
In the same way as what was derived in the previous paragraph, each diagonal 2x2 matrix of $\left(P^{-1}\otimes1\!\!1\right)F{\mathbf{e}_{\rm dc}M\mathbf{I}}F\left(P\otimes1\!\!1\right)$ can be written as
\begin{align}
G_{kk}^{\rm dc}=\frac{g_{k}}{2}V_{k},
\label{appA_Gform_dc}
\end{align}
with
\begin{align}
g_{k}~=~2\sqrt{-\det\left(G_{kk}^{\rm dc}\right)},
\end{align}
and
\begin{align}
V_{k}&=\left(\mathcal{C}^{\rm dc}_x\right)_{k}\sigma_x+\left(\mathcal{C}^{\rm dc}_y\right)_{k}\sigma_y+\left(\mathcal{C}^{\rm dc}_z\right)_{k}\sigma_z,\label{appA_dc_unit}
\end{align}
such that $V_{k}^{\rm dc}$ is unitary with determinant -1. Now the diagonalization of this matrix gives the modification of the eigenvalues of $\mathbf{I}Q\mathbf{I}$, and gives a preferential direction for its eigenvectors, due to the lift of degeneracy. If $\left(\mathcal{C}^{\rm dc}_z\right)_{k}=1$, then $V_{k}=\sigma_z$ is diagonal; if not, all the unitary matrices of the form
\begin{align}
\mathfrak{P}^{\rm dc}_{k}=\frac{V_{k}-\sigma_z}{\left[2-2\left(\mathcal{C}^{\rm dc}_z\right)_{k}\right]^{1/2}}e^{i\varphi^{\rm dc}_{k}}
\label{appA_transfo_inter2}
\end{align}
then allow to diagonalize $V_{k}$, with eigenvalues -1 and +1, where the $\varphi^{\rm dc}_{k}$ can be chosen arbitrarily.\\
\textbf{Eigenvalues:} Formally speaking, this means that the perturbation matrices of the eigenenergies are simply given by:
\begin{align}
B_{\rm dc}\mathfrak{P}^\dagger_kG_{kk}^{\rm dc}\mathfrak{P}_k=-\frac{g_{k}}{2}B_{\rm dc}\sigma_z.
\end{align}
In the body of the text, we have simply noted $|g_{k}B_{\rm dc}|~=~\hbar\delta_k$, such that $g_{k}~=~2\sqrt{-\det\left(G_{kk}^{\rm dc}\right)}$ is the effective gyromagnetic ratio of level $k$.

\textbf{Eigenvectors: }This new basis allows now to rewrite the interaction 2x2 submatrices $U_{kl}^{\rm ac}$ such that 
\begin{align}
\left(U_{kl}^{\rm ac}\right)'=\mathfrak{P}^\dagger_kU_{kl}^{\rm ac}\mathfrak{P}_l,
\label{appA_transfo_inter3}
\end{align} 
where we recall that $U_{kl}^{\rm ac}$ is given by Eq.(\ref{appA_interform_changed2}).\\
 Even if now only a restricted set of unitary transforms are allowed, they still allow to bring the determinant equal to 1. Indeed, as $\det(U_{kl}^{\rm ac})=-1$ we only have to choose the phases  $\varphi_k=-\varphi_l=\pi/2$ (in (\ref{appA_transfo_inter2})) such that $\det\left[\left(U_{kl}^{\rm ac}\right)'\right]=1$.
 
\section{Diagonalization of the $A$ matrix}
\label{sectionAmatrix}
In this appendix, we prove the approximate form of the eigenvalues of the $A$ matrix in the regime $
{(g_s-g_g)^2}/|{g_sg_g}|~<<~1 $ given in Eq.(\ref{eigvalues}).\\
To this extend, let us rewrite the $A$ matrix given by expression (\ref{Amatrix}) in a more suitable basis, that makes it explicit that $u_1$ couples spins of the same sign and $u_2$ couples spins of opposite sign. For that, let us simply swap vectors 2 and 3 to define a new matrix $A'$:
\begin{align}
A'=
\left(
\begin{array}{cccc}
\Delta+\delta_s&e^{i\varphi}\Omega_1&0&e^{i\varphi}\Omega_2\\
e^{-i\varphi}\Omega_1^*&-\Delta+\delta_g&-e^{-i\varphi}\Omega_2&0\\
0&-e^{i\varphi}\Omega_2^*&\Delta-\delta_s&e^{i\varphi}\Omega_1^*\\
e^{-i\varphi}\Omega_2^*&0&e^{-i\varphi}\Omega_1&-\Delta-\delta_g
\end{array}
\right),
\end{align}
where $\Omega_1=u_1\Omega_0$ and $\Omega_2=u_2\Omega_0$, following the notations of Eq.(\ref{defin_unitary_matrices}).\\
We then decompose this matrix in three terms:
\begin{subequations}
\begin{align}
A'=A_{\rm free}+A_{\rm par}+A_{\rm cross},
\end{align}
with
\begin{align}
A_{\rm free}&=
\left(
\begin{array}{cccc}
\Delta+\delta_s&0&0&0\\
0&-\Delta+\delta_g&0&0\\
0&0&\Delta-\delta_s&0\\
0&0&0&-\Delta-\delta_g
\end{array}
\right)\\
A_{\rm par}&=
\left(
\begin{array}{cccc}
0&e^{i\varphi}\Omega_1&0&0\\
e^{-i\varphi}\Omega_1^*&0&0&0\\
0&0&0&e^{i\varphi}\Omega_1^*\\
0&0&e^{-i\varphi}\Omega_1&0
\end{array}
\right)\\
A_{\rm cross}&=
\left(
\begin{array}{cccc}
0&0&0&e^{i\varphi}\Omega_2\\
0&0&-e^{-i\varphi}\Omega_2&0\\
0&-e^{i\varphi}\Omega_2^*&0&0\\
e^{-i\varphi}\Omega_2^*&0&0&0
\end{array}
\right).
\end{align}
\end{subequations}
The interpretation of these matrices is simple: $A_{\rm free}$ characterizes the ion level without interaction, $A_{\rm par}$ the coupling of the levels with no spin flip, and $A_{\rm cross}$ the interaction with a spin flip.\\
To conduct our study, we will first diagonalize the sum $A_{\rm free}+A_{\rm par}$ alone and see in which regime $A_{\rm cross}$ remains unchanged by this diagonalization process.\\
 It is straightforward to see that:
\begin{align}
A_{\rm free}+A_{\rm par}=
\left(
\begin{array}{cccc}
\Delta+\delta_s&e^{i\varphi}\Omega_1&0&0\\
e^{-i\varphi}\Omega_1^*&-\Delta+\delta_g&0&0\\
0&0&\Delta-\delta_s&e^{i\varphi}\Omega_1^*\\
0&0&e^{-i\varphi}\Omega_1&-\Delta-\delta_g
\end{array}
\right),
\end{align}
has the eigenvalues:
\begin{subequations}
\begin{align}
\zeta_1^0&=\frac12\left(\delta_s+\delta_g+\sqrt{(2\Delta+\delta_s-\delta_g)^2+4|\Omega_1|^2}\right)\\
\zeta_2^0&=\frac12\left(\delta_s+\delta_g-\sqrt{(2\Delta+\delta_s-\delta_g)^2+4|\Omega_1|^2}\right)\\
\zeta_3^0&=-\frac12\left(\delta_s+\delta_g-\sqrt{(2\Delta+\delta_g-\delta_s)^2+4|\Omega_1|^2}\right)\\
\zeta_4^0&=-\frac12\left(\delta_s+\delta_g+\sqrt{(2\Delta+\delta_g-\delta_s)^2+4|\Omega_1|^2}\right)
\end{align}
\end{subequations}
and corresponding eigenvectors:
\begin{widetext}
\begin{align}
\mathcal{P}_1=\left(
\begin{array}{cccc}
\cos{\theta_1}&-e^{i(\varphi+\phi_1)}\sin\theta_1&0&0\\
e^{-i(\varphi+\phi_1)}\sin\theta_1&\cos{\theta_1}&0&0\\
0&0&\cos{\theta_2}&-e^{i(\varphi-\phi_1)}\sin\theta_2\\
0&0&e^{-i(\varphi-\phi_1)}\sin\theta_2&\cos{\theta_2}
\end{array}
\right),
\label{passmatrix}
\end{align}
\end{widetext}
where
\begin{subequations}
\begin{align}
\cos\theta_1&=\frac{2\Delta+\delta_s-\delta_g+\sqrt{(2\Delta+\delta_s-\delta_g)^2+4|\Omega_1|^2}}{N_1}\\
\sin\theta_1&=\frac{2|\Omega_1|}{N_1}\\
\cos\theta_2&=\frac{2\Delta+\delta_g-\delta_s+\sqrt{(2\Delta+\delta_g-\delta_s)^2+4|\Omega_1|^2}}{N_2}\\
\sin\theta_2&=\frac{2|\Omega_1|}{N_2},
\end{align}
\label{trigo_rot}
\end{subequations}
$e^{i\phi_1}=u_1/|u_1|$ and $N_1$ and $N_2$ are real normalization factors. As an indication, these four eigenvalues are represented with colored dashed lines in figure \ref{anticross_eig} in the main text. An important point is that the two eigenvalues $\zeta_2^0$ and $\zeta_3^0$ have the possibility to cross, at the point where
\begin{align}
|\Omega_1|^2+4\Delta^2\frac{\delta_s\delta_g}{(\delta_s+\delta_g)^2}=\delta_s\delta_g.\label{crosscond}
\end{align}
Another way of formulating (\ref{crosscond}) is that the two eigenvalues cross for a magnetic field of amplitude 
\begin{align}
B_{\rm dc}=\sqrt{\frac{|\Omega_1|^2}{g_sg_g}+\frac{4\Delta^2}{(g_s+g_g)^2}},
\label{field_for_cross}
\end{align}
where $g_i$ is the effective gyromagnetic ratio of state $i$.

If we now add the cross-interaction term $A_{\rm cross}$, the eigenvalues will be modified. This modification will be noticeable only at the point where the $\zeta_i^0$ values cross, that is at zero field and at the crossing point. In general, $A_{\rm cross}$ is modified according to:
\begin{align}
\mathcal{P}_1^{-1}A_{\rm cross}\mathcal{P}_1=
\left(
\begin{array}{cc}
0&R\\
R^{\dagger}&0
\end{array}
\right),
\end{align}
where 
\begin{align}
R=\Omega_2
\left(
\begin{array}{cc}
e^{i\phi_1}\sin(\theta_2-\theta_1)&e^{i\varphi}\cos(\theta_2-\theta_1)\\
-e^{-i\varphi}\cos(\theta_2-\theta_1)&e^{-i\phi_1}\sin(\theta_2-\theta_1)
\end{array}
\right).
\end{align}
The cross-interaction matrix $A_{\rm cross}$ can then be considered as unchanged by the transformation $\mathcal{P}_1$ in the condition where the ratio between the diagonal terms and the anti-diagonal ones in $R$ is $<<1$. This can simply be re-written as $|\tan(\theta_1-\theta_2)|<<1$, which translates as 
\begin{align}
\frac{|\Omega_1||\delta_s-\delta_g|}{\delta_s\delta_g}<<1,
\end{align}
which at the crossing point can be rewritten, thanks to (\ref{field_for_cross}) like
\begin{align}
Q=\frac{(g_s-g_g)^2}{g_sg_g+\frac{4\Delta^2g_s^2g_g^2}{|\Omega_1|^2(g_s+g_g)^2}}<<1.
\label{condition_eigen}
\end{align}

\begin{figure}[!ht]
\begin{center}
\subfigure[]{\includegraphics[width=4.1cm]{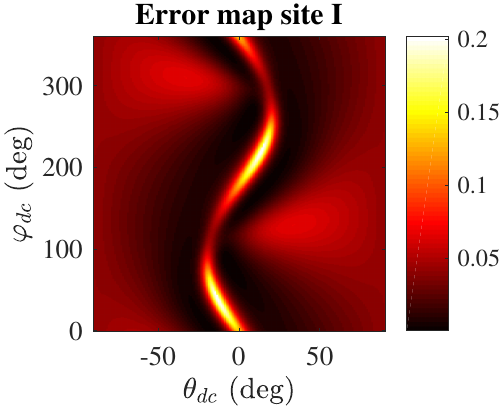}}
\subfigure[]{\includegraphics[width=4.1cm]{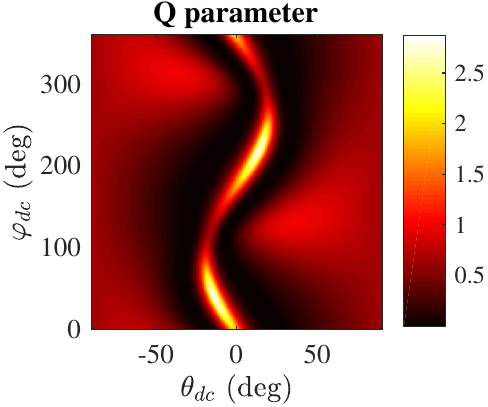}}
\caption{Validity of condition (\ref{condition_eigen}) for site I \cite{Yano91} of our system.(a) Relative error $|\zeta_2-eig_2(A')|/\zeta_2$ between estimated ($\zeta_2$, given by Eq.(\ref{zeta2Appendix})) and actual ($eig_2(A')$) second eigenvalue. A maximal relative error of $\sim$20$\%$ can be seen. (b) Q parameter (\ref{condition_eigen}). As expected, when the condition is not well satisfied (large $Q$), the relative error on the eigenvalue is maximal.}
\label{error_appendix}
\end{center}
\end{figure}

This condition is the one given in the body of the text, with $\Delta=0$. 
When $A_{\rm cross}$ can be considered as unchanged by the $\mathcal{P}_1$ matrix, the eigenvalues of $A'$ are simple to calculate, and can be approximated by:
\begin{subequations}
\begin{align}
\zeta_1&=\frac12\left(\zeta_1^0+\zeta_4^0+\sqrt{(\zeta_1^0-\zeta_4^0)^2+4|\Omega_2|^2}\right)\\
\zeta_2&=\frac12\left(\zeta_2^0+\zeta_3^0+\sqrt{(\zeta_2^0-\zeta_3^0)^2+4|\Omega_2|^2}\right)\label{zeta2Appendix}\\
\zeta_3&=\frac12\left(\zeta_2^0+\zeta_3^0-\sqrt{(\zeta_2^0-\zeta_3^0)^2+4|\Omega_2|^2}\right)\\
\zeta_4&=\frac12\left(\zeta_1^0+\zeta_4^0-\sqrt{(\zeta_1^0-\zeta_4^0)^2+4|\Omega_2|^2}\right).
\end{align}
\label{eigenvalues_A}
\end{subequations}

Notice that we have not performed perturbation theory on $\Omega_2$: the results presented here hold for arbitrary values of $\Omega_2$. However, the reasoning that we have applied here is not symmetric in $\Omega_1$ and $\Omega_2$ as seen with equations (\ref{eigenvalues_A}): the reason for this is that condition (\ref{condition_eigen}) introduces some asymmetry in the role of $\Omega_1$ and $\Omega_2$. Namely, if one first diagonalizes $A_{\rm free}+A_{\rm cross}$, $A_{\rm par}$ will be strongly affected by the transformation.\\
As a sanity check, let us ensure that condition (\ref{condition_eigen}) is relevant for the approximation of the eigenvalues of $A'$. The most sensitive eigenvalues are the one that are the most affected by the avoided crossing, namely $\zeta_2$ and $\zeta_3$. Following this remark, figure \ref{error_appendix}(a) shows the relative distance between $\zeta_2$ given by (\ref{zeta2Appendix}) with the actual second eigenvalue of $A'$: $|\zeta_2-eig_2(A')|/\zeta_2$ as a function of $\varphi_{\rm dc}$ and $\theta_{\rm dc}$, angles of the $\mathbf{B}_{\rm dc}$ field, in the same practical case as in section \ref{sectionInversions}, for site I of $^{151}$Eu$^{3+}$:Y$_2$SiO$_5$ for the ${}^7$F${}_0$ $\ket{\pm3/2}_G\leftrightarrow\ket{\pm1/2}_G$ nuclear spin transition. On the other hand, if one plots the $Q$ parameter given by Eq.(\ref{condition_eigen}), one obtains the dependency shown on fig.\ref{error_appendix}(b). The similarity of the two plots clearly validates the condition that we have found.

\section{Expression of the propagator at low field}
\label{appendix_propag_low}
In this appendix, we derive the expression of the propagator at low field. In order to simplify the calculations, we place ourselves in the resonant case ($\Delta=0$).\\
Given that the Schr\"odinger equation simply reduces to (\ref{eqpsi1}), the propagator reads:
\begin{align}
U_{\rm prop}=e^{iAt/2}=P\cdot e^{iDt/2}\cdot P^{-1},
\label{propag_zero}
\end{align}
where $P$ is the transfer matrix that diagonalizes $A$ to $D$:
\begin{align}
D=\left(
\begin{array}{cccc}
\zeta_1&0&0&0\\
0&\zeta_3&0&0\\
0&0&\zeta_2&0\\
0&0&0&\zeta_4
\end{array}
\right)=
\left(
\begin{array}{cccc}
\zeta_1&0&0&0\\
0&\zeta_3&0&0\\
0&0&-\zeta_3&0\\
0&0&0&-\zeta_1
\end{array}
\right).
\end{align}
We remind that the eigenvalues obtained in Appendix \ref{sectionAmatrix} are the ones of $A'$, for which vectors 2 and 3 were exchanged (we note the corresponding flip matrix $\mathcal{P}_{23}$), hence a change in the indexes of the eigenvalues in $D$. Then, at zero detuning, we simply have $\zeta_2=-\zeta_3$ and $\zeta_4=-\zeta_1$. The transfer matrix is then given by 
\begin{align}
P=\mathcal{P}_{23}\mathcal{P}_1\mathcal{P}_2,
\label{basischangeA}
\end{align}
where $\mathcal{P}_1$ is given in Eq.(\ref{passmatrix}) and $\mathcal{P}_2$ is the transfer matrix from the base in which $A_{\rm free}+A_{\rm par}$ is diagonal to the base where $A'$ is diagonal. In the case $\Delta=0$, it simply reads
\begin{widetext}
\begin{align}
\mathcal{P}_2=\left(
\begin{array}{cccc}
\cos\theta_4&0&0&-e^{i(\phi_2+\varphi)}\sin\theta_4\\
0&\cos\theta_3&-e^{i(\phi_2-\varphi)}\sin\theta_3&0\\
0&e^{-i(\phi_2-\varphi)}\sin\theta_3&\cos\theta_3&0\\
e^{-i(\phi_2+\varphi)}\sin\theta_4&0&0&\cos\theta_4
\end{array}
\right),
\label{passmatrix2}
\end{align}
\end{widetext}

where
\begin{subequations}
\begin{align}
\cos\theta_3&=\frac{-\zeta_2^0+\sqrt{\left(\zeta_2^0\right)^2+|\Omega_2|^2}}{N_3}\\
\sin\theta_3&=\frac{|\Omega_2|}{N_3}\\
\cos\theta_4&=\frac{\zeta_1^0+\sqrt{\left(\zeta_1^0\right)^2+|\Omega_2|^2}}{N_4}\\
\sin\theta_4&=\frac{|\Omega_2|}{N_4},
\end{align}
\label{trigo_rot2}
\end{subequations}
and $e^{i\phi_2}=u_2/|u_2|$.
Now that we have the explicit expression of the matrices, we can write the propagator. However, its general expression without approximation is not so trivial. As we are interested in its expression at low fields, we can consider that $\epsilon=\frac{\delta_g+\delta_s}{2\Omega_0}<<1$.
Under this approximation,
\begin{subequations}
\begin{align}
\zeta_1&\simeq\Omega_0+\epsilon|\Omega_1|\\
\zeta_3&\simeq-\Omega_0+\epsilon|\Omega_1|,
\end{align}
\end{subequations}

\begin{subequations}
\begin{align}
\cos\theta_1&\simeq \cos\theta_2\simeq \sin\theta_1\simeq \sin\theta_2\simeq 1/\sqrt{2},\\
\cos\theta_3&\simeq \cos\theta_4\simeq \frac{\Omega_0+|\Omega_1|}{\sqrt{2\Omega_0(\Omega_0+|\Omega_1|)}}:=c_0=\sqrt{\frac{1}{2}}\sqrt{1+|u_1|}\\
\sin\theta_3&\simeq \sin\theta_4\simeq \frac{|\Omega_2|}{\sqrt{2\Omega_0(\Omega_0+|\Omega_1|)}}:=s_0=\sqrt{\frac{1}{2}}\sqrt{1-|u_1|}.
\end{align}
\end{subequations}
The transfer matrix then simply reads:
\begin{widetext}
\begin{align}
P\simeq
\frac{1}{\sqrt{2}}\left(
\begin{array}{cccc}
c_0&-c_0e^{i(\phi_1+\varphi)}&s_0e^{i(\phi_1+\phi_2)}&-s_0e^{i(\phi_2+\varphi)}\\
-s_0e^{-i(\phi_1+\phi_2)}&s_0e^{-i(\phi_2-\varphi)}&c_0&-c_0e^{-i(\phi_1-\varphi)}\\
c_0e^{-i(\phi_1+\varphi)}&c_0&-s_0e^{i(\phi_2-\varphi)}&-s_0e^{i(\phi_2-\phi_1)}\\
s_0e^{-i(\phi_2+\varphi)}&s_0e^{-i(\phi_2-\phi_1)}&c_0e^{i(\phi_1-\varphi)}&c_0
\end{array}
\right),
\label{matpass_approx}
\end{align}
and
\begin{align}
\exp\left(\frac{i}{2}Dt\right)\simeq
\left(
\begin{array}{cccc}
e^{\frac{i}{2}\Omega_0t}e^{\frac{i}{2}\epsilon|\Omega_1|t}&0&0&0\\
0&e^{-\frac{i}{2}\Omega_0t}e^{\frac{i}{2}\epsilon|\Omega_1|t}&0&0\\
0&0&e^{\frac{i}{2}\Omega_0t}e^{-\frac{i}{2}\epsilon|\Omega_1|t}&0\\
0&0&0&e^{-\frac{i}{2}\Omega_0t}e^{-\frac{i}{2}\epsilon|\Omega_1|t}
\end{array}
\right).
\label{expdiag_approx}
\end{align}
\end{widetext}
Injecting (\ref{matpass_approx}) and (\ref{expdiag_approx}) into (\ref{basischangeA}) and (\ref{propag_zero}), we get the expression of the propagator:
\begin{align}
U_{\rm prop}(t)\simeq \cos\left(\frac{\epsilon|\Omega_1|t}{2}\right)U_0(t)+\sin\left(\frac{\epsilon|\Omega_1|t}{2}\right)U_{\rm pert}(t),
\label{propag_tot}
\end{align}
with $U_0(t)$ the propagator at zero field
\begin{align}
U_{0}(t)=\cos\left(\frac{\Omega_0t}{2}\right)1\!\!1+i~{\sin\left(\frac{\Omega_0t}{2}\right)}
\left(
\begin{array}{cc}
0&e^{i\varphi}U\\
e^{-i\varphi}U^\dagger &0
\end{array}
\right)
\label{spinPropNoBAppen}
\end{align}
 and
\begin{align}
&U_{\rm pert}(t)=i\cos\left(\frac{\Omega_0t}{2}\right)
\left(
\begin{array}{cccc}
|u_1|&-u_2e^{i\phi_1}&0&0\\
-u_2^*e^{-i\phi_1}&-|u_1|&0&0\\
0&0&|u_1|&u_2e^{-i\phi_1}\\
0&0&u_2^*e^{i\phi_1}&-|u_1|
\end{array}
\right)\nonumber\\
&+\sin\left(\frac{\Omega_0t}{2}\right)
\left(
\begin{array}{cccc}
0&0&-e^{i(\varphi+\phi_1)}&0\\
0&0&0&e^{i(\varphi-\phi_1)}\\
-e^{-i(\varphi+\phi_1)}&0&0&0\\
0&e^{-i(\varphi-\phi_1)}&0&0
\end{array}
\right).
\end{align}

\bibliographystyle{apsrev4-2}
\bibliography{RefPD.bib}

\end{document}